\documentclass[fleqn,usenatbib]{mnras}

\usepackage{newtxtext,newtxmath}
\usepackage{lscape}

\usepackage[T1]{fontenc}
\usepackage{ae,aecompl}

\newbox\grsign \setbox\grsign=\hbox{$>$} \newdimen\grdimen \grdimen=\ht\grsign
\newbox\simlessbox \newbox\simgreatbox
\setbox\simgreatbox=\hbox{\raise.5ex\hbox{$>$}\llap
     {\lower.5ex\hbox{$\sim$}}}\ht1=\grdimen\dp1=0pt
\setbox\simlessbox=\hbox{\raise.5ex\hbox{$<$}\llap
     {\lower.5ex\hbox{$\sim$}}}\ht2=\grdimen\dp2=0pt

\newbox\simppropto
\setbox\simppropto=\hbox{\raise.5ex\hbox{$\sim$}\llap
     {\lower.5ex\hbox{$\propto$}}}\ht2=\grdimen\dp2=0pt

\usepackage{color}
\usepackage[dvipsnames]{xcolor}

\usepackage{textcomp}
\def\pm{\textpm}

\usepackage{graphicx}	

\title[Neutron capture elements in the Galactic disc]{Neutron-capture elements record the ordered chemical evolution of the disc over time }

\author[D. Horta et al.]{
Danny Horta$^{1,2}$,\thanks{E-mail: D.HortaDarrington@2018.ljmu.ac.uk}
Melissa K. Ness$^{3,4}$,
Jan Rybizki$^{5}$,
Ricardo P. Schiavon$^{1}$, 
Sven Buder$^{6,7}$
\\
$^{1}$ Astrophysics Research Institute, Brownlow Hill,  Liverpool, L3 5RF, UK\\
$^{2}$ School of Mathematics and Physics, The University of Queensland, St. Lucia, QLD 4072, Australia\\
$^{3}$ Department of Astronomy, Columbia University, Pupin Physics Laboratories, New York, NY 10027, USA\\
$^{4}$ Center for Computational Astrophysics, Flatiron Institute, 162 Fifth Avenue, New York, NY 10010, USA\\
$^{5}$ Max-Planck-Institut f\"ur Astronomie, K\"onigstuhl 17, 69117 Heidelberg, Germany\\
$^{6}$ Research School of Astronomy $\&$ Astrophysics, Australian National University, Canberra, ACT 2611, Australia\\
$^{7}$ ARC Centre of Excellence for All Sky Astrophysics in 3 Dimensions (ASTRO 3D), Australia
}
\date{Accepted XXX. Received YYY; in original form ZZZ}
\pubyear{20XX}
\begin{document}
\label{firstpage}
\pagerange{\pageref{firstpage}----\pageref{lastpage}}
\maketitle
\begin{abstract}

An ensemble of chemical abundances probing different nucleosynthetic channels can be leveraged to build a comprehensive understanding of the chemical and structural evolution of the Galaxy. Using GALAH DR3 data, we seek to trace the enrichment by the supernovae Ia, supernovae II,  asymptotic giant branch stars, and neutron-star mergers and/or collapsars nucleosynthetic sources by studying the [Fe/H], [$\alpha$/Fe], [Ba/Fe], and [Eu/Fe] chemical compositions of $\sim$50,000 red giant stars, respectively. Employing small [Fe/H]-[$\alpha$/Fe] cells, which serve as an effective reference-frame of supernovae contributions, we characterise the abundance-age profiles for [Ba/Fe] and [Eu/Fe]. Our results disclose that these age-abundance relations vary across the [Fe/H]-[$\alpha$/Fe] plane. Within cells, we find negative age-[Ba/Fe] relations and flat age-[Eu/Fe] relations. Across cells, we see the slope of the age-[Ba/Fe] relations evolve smoothly and the [Eu/Fe] relations vary in amplitude. We subsequently model our empirical findings in a theoretical setting using the flexible Chempy Galactic chemical evolution (GCE) code, using the mean [Fe/H], [Mg/Fe], [Ba/Fe], and age values for stellar populations binned in [Fe/H], [Mg/Fe], and age space. We find that within a one-zone framework, an ensemble of GCE model parameters vary to explain the data. Using present day orbits from \textit{Gaia} EDR3 measurements we infer that the GCE model parameters, which set the observed chemical abundance distributions, vary systematically across mean orbital radii. Under our modelling assumptions, the observed chemical abundances are consistent with a small gradient in the high mass end of the initial mass function (IMF) across the disc, where the IMF is more top heavy towards the inner disc and more bottom heavy in the outer disc. 

\end{abstract}


\section{Introduction}
\label{Introduction}

Chemical abundances of stars encode properties of the environment in which they form and can be used to characterise the formation of the Milky Way. Beginning with the pioneering studies of the Sun and nearby stars \citep[][]{Payne1926, Russell1929}, the pursuit of a complete element genesis remains as a fundamental question, relevant to both astrophysics and nuclear physics (\citealp[e.g.,][]{Fowler1955,Burbidge1957,Wallerstein1997}). Significant progress has been made over the last half-century, and we can classify most elements into a small subset of primary groups, based on the their dominant formation nucleosynthetic channels. Broadly speaking, elements can be organised into the $\alpha$, odd-Z, iron-peak, and neutron capture groups, and into further subgroups within each. The analysis of these different elemental classifications in different chemical planes can subsequently be used to connect the relative conditions in which a stellar population is formed to the way the Galaxy has formed and evolved over time. 

Progress in our understanding of chemical element distributions in stars in the Milky Way has been driven by large spectroscopic stellar surveys. These include \citealp[APOGEE:][]{Majewski2017}, \citealp[ GALAH:][]{Martell2017,Buder2020}, \citealp[LAMOST:][]{Zhao2012}, \citealp[RAVE:][]{Steinmetz2020}, \citealp[SEGUE:][]{Segue2009}, \citealp[$Gaia$-ESO:][]{Gilmore2012}, amongst others. The ability to determine precise chemical abundances from spectra for $>$ $10^5-10^6$ stars in the Galaxy is providing a wealth of data to be explored. Chemical abundance measurements, when coupled with precise phase space information from the \textit{Gaia} mission \citep[][]{Gaia2020}, and stellar ages \citep[e.g.,][]{Ness2016,Leung2019,Sharma2020, Hayden2020}, permit a full chemo-dynamical-age exploration of stellar populations, especially in the most massive Galactic stellar component: the Milky Way's disc (\citealp[e.g.,][]{Gratton1996,Fuhrmann1998,Fuhrmann2004,Adibekyan2011,Bovy2012,bovy2012b,Minchev2013,Bensby2014,Minchev2014,Hayden2015,Martig2016,Ness2016,Mackereth2017,Minchev2017,Antoja2018,Ciuca2018,Katz2018,Blancato2019,Bland2019,Bovy2019,Buder2019,Ness2019,Mackereth2019_disc,Hayden2020,Ciuca2021,Yuxi2021}) \footnote{This is of course an incomplete list (and highlights only some of the most recent work). For a comprehensive review on studies undertaken on unravelling the Milky Way disc, we refer the reader to Section 5 from \citet{Bland2016} and references therein.}. 

Chemo-dynamical studies of the Galactic disc have revealed its multifaceted nature. Chemically, the disc is dominated in character by a bimodality observed in the [Fe/H]-[$\alpha$/Fe] plane, that correlates with its orbital and spatial distribution properties. The origin of the $\alpha$-bimodality is uncertain, and can be explained under different mechanisms seen in simulations (\citealp[e.g.,][]{ Mackereth2018,Buck2019,Grand2018, Grand2020,vintergata2020, K2021}). Recent observations suggest that it is perhaps a not uncommon phenomenon \citep{SN2021}, although some theoretical results propose the contrary \citep[][]{Mackereth2018}. From a nucleosynthetic standpoint, the [Fe/H]-[$\alpha$/Fe] plane represents two source production sites. The $\alpha$-elements are understood to be produced primarily in massive star progenitors via core-collapse supernovae (\citealp[SNII; e.g.,][]{Timmes1995,Kobayashi2020}) and  iron-peak elements are produced via white dwarf detonations (SNIa). The [Fe/H]-[$\alpha$/Fe] plane has served as a critical diagnostic of the evolutionary history of various components of our Galaxy. Here our goal is to take a next step in understanding the chemical signatures in the Milky Way disc. We set out to characterise and interpret the abundance-age relations of the little studied neutron capture elements in Galactic disc populations. We do this using the latest data release from the GALAH survey \citep[DR3;][]{Buder2020}. We reason that exploration of abundance relations as a function of age is pertinent to addressing  questions of the formation, and evolution, of the Galactic disc which occur over time. We wish to undertake as controlled an experiment as possible. Therefore, we examine the distribution of the neutron-capture abundances conditioned on other variables; i.e. keeping other chemical properties of the stars fixed. Specifically, we examine the neutron-capture element abundances over time in small chemical cells  (\citealp[or MAPS; e.g.,][]{Bovy2012,Mackereth2020,Yuxi2021}) across the [Fe/H]-[$\alpha$/Fe] plane. This grid across the well studied  [Fe/H]-[$\alpha$/Fe] plane (our chemical cells) allows us to control for the dependency on the contribution from two (dominant) types of supernovae within chemical cells (namely, the main drivers of the chemical evolution of the Galaxy) when interpreting the behaviour of the neutron-capture elements at different ages. We utilise this approach to study Barium (Ba) and Europium (Eu) as representative  $s$- and $r$-process neutron capture elements, respectively. These neutron capture elements are of particular interest, as recent studies have suggested that the $r$-process elements may originate in significant amounts from exotic neutron star mergers (\citealp[e.g.,][]{Metzger2010,Kasen2017,Siegel2017,Tanaka2017,Siegel2019,Watson2019}), whilst the $s$-process are important tracers of the contribution from stellar winds of evolved asymptotic giant branch (AGB) stars (\citealp[e.g.,][]{Sneden2008,Kobayashi2020b}).

The empirical characterisation of abundances over time in a reference plane of supernovae contributions, as enabled by large survey data, is the first step of our study. The next is interpretation of the trends identified within a theoretical framework. The ensemble data that we examine are compared to the predictions of Galactic Chemical Evolution models (GCE), which describe the star formation and chemical enrichment of the interstellar medium in terms of a set of mass conservation equations. This technique has been adopted extensively (\citealp[e.g.,][]{Tinsley1980,Matteucci1989,Prantzos1993,Timmes1995,Pagel1997,Chiappini1997,McWilliam1997,Kobayashi2000,Matteucci2016,Kobayashi2020, matteucci2021modelling}). It has also led to the development of sophisticated and publicly available software packages (\citealp[e.g.,][]{Andrews2017,Rybizki2017, Philcox2019, Johnson2021}), making this method  accessible to many studies aimed at interpreting the chemical evolution of the Milky Way.

Chemical evolution modelling is challenging. There are many uncertainties and even the Solar abundances can not be reproduced under the most flexible of approaches, indicative of inaccuracies in the yields \citep[e.g.][]{Rybizki2017, Blancato2019}. Detailed and complex multi-zone frameworks with prescriptions of migration offer perhaps the most theoretically complete approach to reconciling theory with observations (\citealp[e.g.][]{Johnson2021}). Yet, under fixed environmental model parameters or simple star formation histories, these do not reproduce the observed distribution of the Galactic disc in the [Fe/H]-[$\alpha$/Fe] plane. Only multi-zone prescriptions with complex star formation histories can reproduce mean chemical trends in the Milky Way's disc and bulge. For example, those that include gas infall events, or optimise for star formation delay times and a prescription of bursty star formation \citep[e.g.][]{Matteucci2019,Lian2020,Matteucci2020,Sharma2020,Spitoni2021}. 

Here, we adopt a flexible one-zone modelling approach using the Chempy GCE software \citep[e.g., $Chempy$:][]{Rybizki2017}. A flexible framework is core to our goals, which are to identify what GCE model parameters the data are consistent with \citep[e.g.][]{Gutcke2019}. Understanding what relations exist both elucidates insight into the source genesis of the elements and quantifies how environmental parameters might vary over the Galaxy's spatial and temporal architecture. The chemical abundances that we model are [Fe/H], [Mg/Fe] and [Ba/Fe], at different ages. The parameters that we set out to constrain are the number of type Ia progenitors, the peak in the Star Formation Rate (SFR), the Star Formation Efficiency (SFE) and the high-mass slope of the Initial Mass Function (IMF). By modelling our abundance data for three representative and dominant nucleosyntheic channels (namely, SNIa, SNII, and AGB winds) within a flexible GCE framework, and then exploring the GCE model parameters over time, we hope to learn about the chemical evolution history of the Galaxy. This in turn can serve as a test bed for current galaxy formation models.

Characterising the star formation environment across the Milky Way disc connects directly to the broader extra-galactic context. Numerous theoretical predictions and interpretations of extra-galactic observations rely on  stellar population parameters. The SFR and SFE have been extensively studied. Variation between and within galaxies is reported for these parameters (\citealp[e.g.][]{Schaefer2017,Krumholz2018,Lopez2018,Medling2018,Rowlands2018,Spindler2018}). The IMF is less well directly constrained by the data, but is fundamental to understanding the mass assembly of galaxies. First introduced by \citet{Salpeter1955}, the IMF provides a convenient way of parameterising the relative numbers of stars as a function of their mass. Its functional form is input as an assumption to many astrophysical works. Until recent years, studies have favoured the scenario of a universal IMF (\citealp[e.g.,][]{Salpeter1955,Miller1979, Scalo1986,Hawkins1988,Kroupa1993, Scalo1998, Kroupa2001, Kroupa2002, Reid2002, Bastian2010, Kaliari2013}), despite the contention of its exact functional form. However, in the last few decades there has been accumulating evidence for variations in the IMF, both in external galaxies (\citealp[e.g.,][]{Hoversten2008,Treu2010,Conroy2012,Cappellari2012,Cappellari2013,Ferreras2013,Lyubenova2016,Davis2017,Vandokkum2017, Parikh2018,Dominguez2019}), and in the Milky Way (\citealp[e.g.,][]{Ballero2007,Brandner2008,Espinoza2009, Lu2013, Hallakoun2020, Griffith2021}), suggesting that the IMF varies in a complicated manner with environment or alternative interpretations of extragalactic observations must be explored \citep[e.g.,][]{Guszejnov2019}. 

The paper is organised as follows: Section \ref{data} reports the data employed in this study and the main selection criteria used to determine our parent working sample. Section~\ref{method} describes the methods used to analyse and model the data. It also includes a description of the GCE models utilised to model the observational data. In Section~\ref{results}, we present the empirical abundance-age relations for Ba and Eu within chemical cells (Section~\ref{sec_cells}) and their correlation with orbital radius. In Section~\ref{sec_modelling} we show the results from modelling our empirical findings using GCE models. We then discuss the implications of our results in the context of previous work in Section~\ref{discussion}, and finalise by presenting our conclusions in Section~\ref{conclusion}.

\section{Data} 
\label{data}

We use data from the second edition of the third data release (DR3) of the GALactic Archaeology with HERMES (GALAH) survey (\citealp{Martell2017,Buder2020}). GALAH measures chemical element abundances and radial velocities using high-resolution ($R$ $\sim$ 28,000) optical spectra collected by the HERMES spectrograph \citep{Sheinis2015} with 2dF fibre positioning system \citep[][]{Lewis2002} mounted on the 3.9-metre Anglo-Australian Telescope at Siding Spring Observatory, Australia. GALAH DR3 was constructed by combining the 2MASS \citep[][]{Skrutskie2006} catalogue of infrared photometry with the UCAC4 \citep[][]{Zacharias2013} proper motion catalogue. It contains data for four main projects (GALAH-main, GALAH-faint, K2-HERMES, and TESS-HERMES), each with its own selection function. The main GALAH survey comprises 74$\%$ of the GALAH DR3 catalogue. GALAH-main targets are all stars with 12.0 < $V$ < 14.0, $\delta$ < +10$^{\circ}$, and |$b$| > 10$^{\circ}$ in regions of the sky that have a minimum of 400 targets in $\pi$ squared degrees (the 2dF field of view), and all stars in the same sky region with 9.0 < $V$ < 12.0 and at least 200 stars per 2dF field of view. K2-HERMES (16$\%$ of the survey) aims to observe targets from the NASA K2 mission, and observes stars in the range 10 < $V$ < 13 or 13 < $V$ < 14
with ($J$ $-$ $K_{s}$) $>$ 0.5. TESS-HERMES observations (7$\%$ of the survey) focused
on stars in the TESS apparent magnitude range (10.0 $<$ $V$ $<$ 13.1) and Southern viewing zone, within 12 degrees of the
Southern ecliptic pole. The remaining $\sim$3$\%$ consists of targets focused on open/globular clusters. 

All data from HERMES were reduced with the \texttt{iraf} pipeline \citep[see][for details]{kos2017}, and is analysed with the Spectroscopy Made Easy (\texttt{SME}) software \citep[][]{Piskunov2016}, which performs spectrum synthesis for 1D stellar atmosphere models. The \texttt{MARCS} theoretical 1D hydrostatic models \citep[][]{Gustafsson2008} are used with spherically symmetric stellar atmosphere models for log$g$ $\leq$ 3.5 and plane parallel models otherwise. For its radiative transfer calculations, SME can take into account departures from local thermodynamic equilibrium (LTE) through grids of departure coefficients. These are known to be significant for several elements like Lithium \citep{Lind2009}. 1D Non-LTE departure coefficients for the eleven 11 elements H, Li, C, O, Na, Mg, Al, Si, K, Ca, Ti, Mn, Fe, and Ba were taken into account \citep[for more details see][]{Amarsi2020}.

In addition to the main GALAH DR3 catalogue, we use the value added catalogues of stellar ages and orbital information. Here, stellar ages were determined using the Bayesian Stellar Parameter Estimation code \citep[BSTEP:][]{Sharma2018}. BSTEP provides a Bayesian estimate of intrinsic stellar parameters from observed parameters by making use of stellar isochrones, adopting a flat prior on age and metallicity (see \citet{Sharma2018} and \citet{Buder2020} for details). Conversely, the orbital information was determined by combining stellar distances (from BSTEP where possible and from prior-informed values by \citet{Bailer2018} when not) and 5D astrometric information from \textit{Gaia} EDR3 \citep[][]{Gaia2020}. Here, orbits of stars were estimated employing the publicly available \texttt{galpy} (\citealp[][]{Galpy2015,Galpy2018}) software package assuming a \citet{McMillan2017} Milky Way potential, and adopting a solar radius of 8.21 kpc (consistent with the latest measurement of 8.178 ± 0.013 (stat) ± 0.022 (sys) kpc \citealp{Gravity2019}), a solar vertical position above the Galactic plane of 25 pc \citep[][]{Juric2008}, and a circular velocity at this radius of 233.1 kms$^{-1}$. 

For more information on the targets, data reduction methods, and value added catalogues used to obtain the final GALAH DR3 catalogue, see \citet{Buder2020}.

\subsection{Selecting a parent disc sample}
For this work, we concern ourselves with a parent sample comprised of Red Giant Stars (RGB) from the second edition of the GALAH Data Release 3 catalogue of stellar parameters and abundances. Stars formed part of the parent sample if they satisfied the following selection criteria:

\begin{itemize}
    \item 4000 < T$_{\mathrm{eff}}$ < 6000 (K) and log$g$ < 3.6, 
    \item \texttt{flag\_sp} = 0 and  \texttt{flag\_X\_Fe} = 0 (for Mg, Fe, Ba, and Eu), 
    \item \texttt{ruwe} < 1.4 and \texttt{parallax$_{-}$over$_{-}$error} > 5,
    \item \texttt{snr$_{-}$c3$_{-}$iraf} > 50.
\end{itemize}

    The T$_{\mathrm{eff}}$ and log$g$ cuts were employed to remove any stars with temperature values too hot/cold for the abundances to be affected by, and to select giant stars, respectively. The \texttt{flag-sp} and \texttt{flag-X-Fe} flags were set to obtain a sample of stars for which reliable abundances were determined over all, and specifically for the elements of upmost importance in this study (namely, Mg, Fe, Ba, and Eu). The \texttt{ruwe} and \texttt{parallax$_{-}$over$_{-}$error} cuts were employed to determine stars with reliable orbital parameters. The \texttt{snr$_{-}$c3$_{-}$iraf} cut was performed to only select stars with a good signal-to-noise ratio in GALAH. 

In addition to the quality cuts, we further remove any stars in the vicinity of the Large and Small Magellanic (L/SMC) clouds following the methodology in \citet{Martell2020}. To do so, we removed stars with:
\begin{itemize}
    \item 73 < $\alpha$ < 83 ($^{\circ}$), --73 < $\delta$ < --63 ($^{\circ}$), 0.3 < pm$_{\alpha}$ < 3.3 (mas/yr), --1.25 < pm$_{\delta}$ < 1.75 (mas/yr), and $v_{\mathrm{helio}}$ > 215 (kms$^{-1}$) for potential LMC star candidates.
    \item  7 < $\alpha$ < 17 ($^{\circ}$), --79 < $\delta$ < --69 ($^{\circ}$), --1.35 < pm$_{\alpha}$ < 2.35 (mas/yr), --2.7 < pm$_{\delta}$ < 0.3 (mas/yr), $v_{\mathrm{helio}}$ > 80 (kms$^{-1}$), and  $\varpi$ < 0.08 (arcsec) for potential SMC star candidates.
\end{itemize}
 This removed a total of 137 LMC and 87 SMC potential star candidates, respectively. The resulting parent sample resulted in a total of 56,927 stars.

\begin{figure}
    \centering
    \includegraphics[width=\columnwidth]{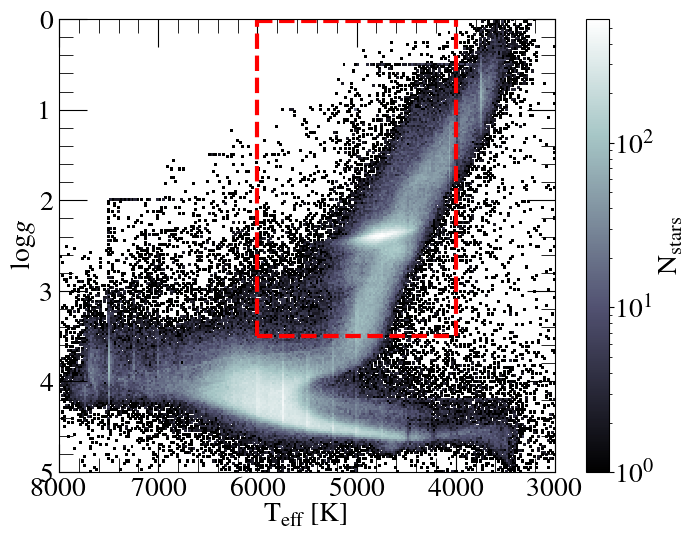}
    \caption{2D histogram of the full GALAH DR3 data in the spectroscopist's Hertzprung-Russell diagram. The red dashed lines indicate the region containing the parent RGB sample of 52,057 disc stars employed in this study.}
    \label{hr}
\end{figure}

\begin{figure}
    \centering
    \includegraphics[width=\columnwidth]{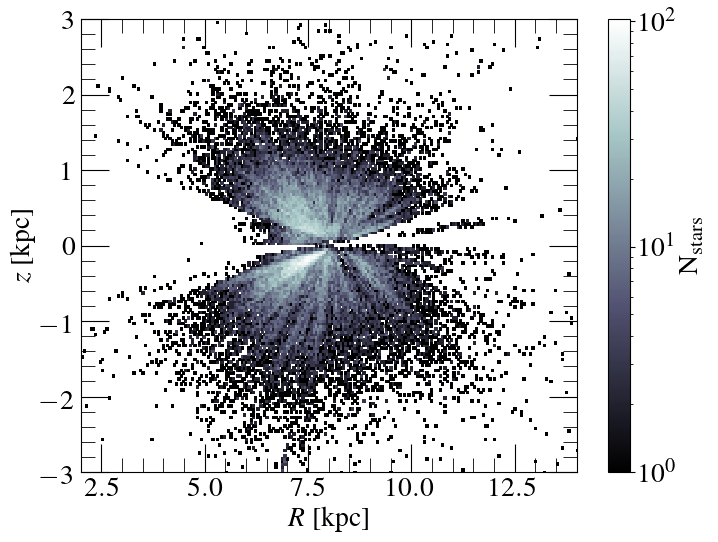}
    \caption{2D histogram illustrating the spatial distribution of the parent sample (i.e., stars in the red box in Fig~\ref{hr}) in the $R$-$z$ plane. The sample is primarily contained within |$z$| $\lesssim$ 2 kpc and 4 $\lesssim$ $R$ $\lesssim$ 12 kpc.}
    \label{Rz}
\end{figure}

\begin{figure*}
    \centering
    \includegraphics[width=\textwidth]{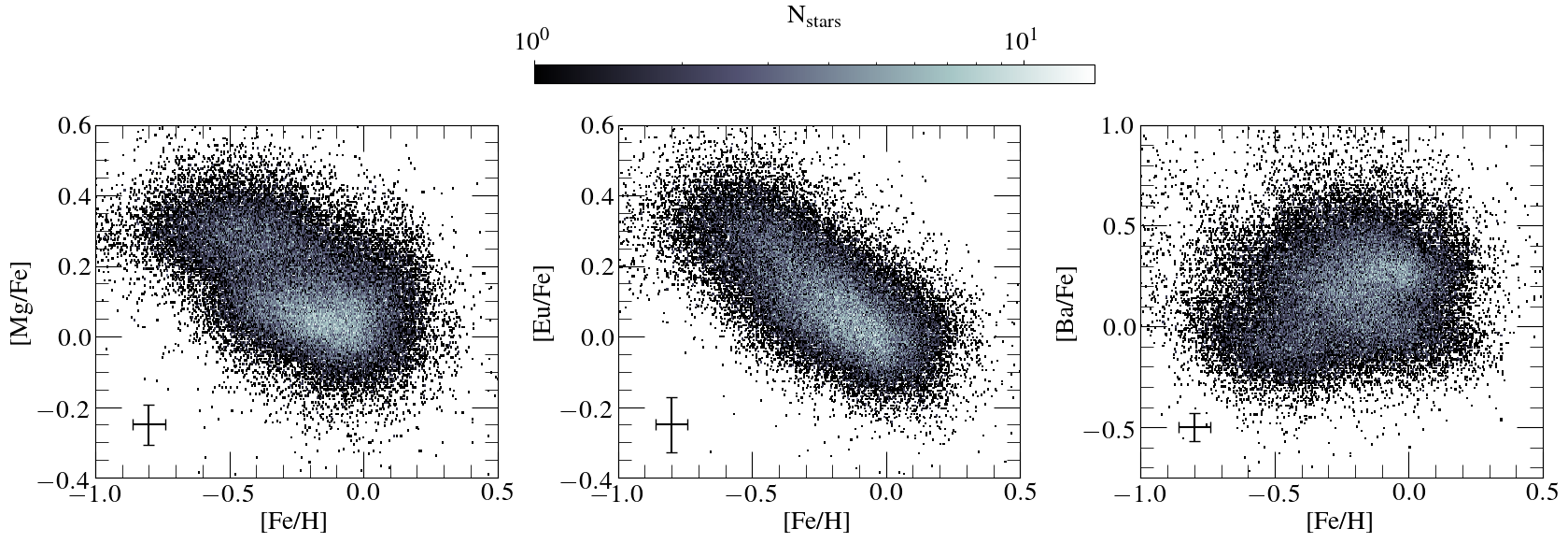}
    \caption{2D density distribution of our parent sample in the [Mg/Fe] (left), [Eu/Fe] (centre), and [Ba/Fe] (right) vs [Fe/H] planes.}
    \label{abundances_sample}
\end{figure*}

\begin{figure}
    \centering
    \includegraphics[width=\columnwidth]{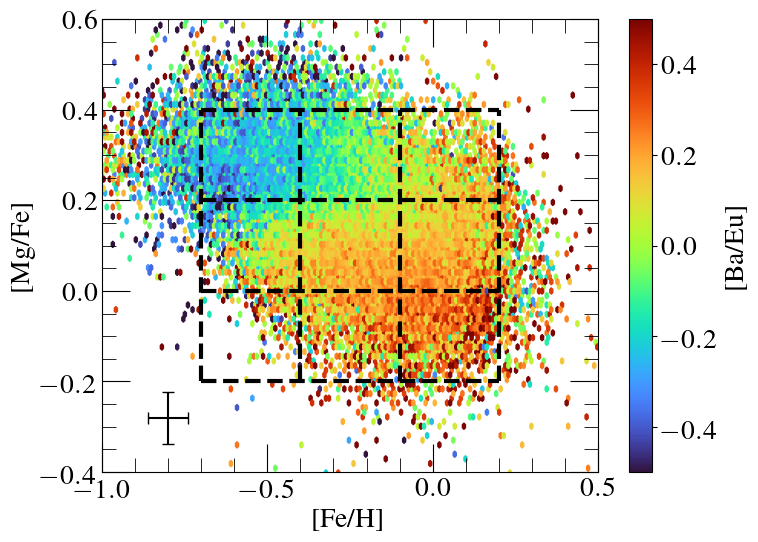}
    \caption{Binned distribution of the parent sample in the [Fe/H]-[Mg/Fe] plane, colour coded by [Ba/Eu]. Here, high [Ba/Eu] indicates stars that are more $s$-process enhanced relative to the $r$-process, and vice-versa. Dashed black lines indicate the boundaries of the grid of 9 cells into which we divide our parent sample in this plane and model their chemical evolution in Section~\ref{sec_modelling}. As can be seen, the low-$\alpha$ disc is more $s$-process enhanced, whilst the high-$\alpha$ disc is more $r$-process enhanced. }
    \label{mgfe}
\end{figure}

Since the aim of this work is to study the $s$- and $r$- process abundance profiles as a function of age for Galactic disc populations, we impose a further metallicity cut of [Fe/H] > --1 and an eccentricity cut of $e$ < 0.4 in order to remove any contamination from metal-poor (primarily halo) populations, and stars on highly eccentric orbits, respectively, as these populations are likely to not belong to the disc of the Galaxy. These further selection criteria remove a total of 4,870 stars. The resulting parent working sample is illustrated in the spectroscopist's HR diagram in Fig~\ref{hr} and spatially in Fig~\ref{Rz}. We also display the parent sample chemically in the [Mg/Fe], [Eu/Fe], and [Ba/Fe] vs [Fe/H] planes in Fig~\ref{abundances_sample}, where we further show the [Ba/Eu] of each star in Fig~\ref{mgfe}. The final parent sample is comprised of 52,057 stars.

\section{Methods}
\label{method}

Our analysis approach is two-pronged. We first undertake an empirical examination of the neutron-capture abundance trends with age in the disc of the Milky Way. Second, we use a chemical evolution modelling framework to understand what parameters of the GCE model are consistent with the data. Specifically, for our empirical analysis, we examine the age-[Ba/Fe] trends and age-[Eu/Fe] trends using chemical cells (or MAPs: \citealp[][]{Bovy2012,Mackereth2020,Yuxi2021}) defined in the [Fe/H]-[$\alpha$/Fe] plane (where we use Mg as the $\alpha$-element), to probe the $s$- and $r$-process nucleosynthetic channels as a function of age across a SNII-SNIa reference frame. Following, we utilise the mean ([Fe/H], [Mg/Fe], [Ba/Fe], age) value for stellar populations binned across [Fe/H], [Mg/Fe], and age space as representative stars and undertake chemical evolution modelling, with the aim of learning what parameters in the GCE model are consistent with the data.

A core and reasonable assumption in this work, that motivates the chemical cells framework of our approach, is that the three measurements of ([Fe/H],[Mg/Fe],age) provide a link to the birth radius of stars in the Milky Way (\citealp[e.g.,][]{Frankel2018,Frankel2019,Frankel2020,Minchev2018,Feltzing2020,Ness2019,Ness2021,Yuxi2021}). This assumption has additional empirical support not only in observed radial gradients in [Fe/H] today, but also in analyses of stars of common ([Fe/H],[X/Fe]), which shows that stars clustered using their chemical abundances group in present day spatial properties and age (\citealp[][]{Ratcliffe2020}). Cosmological simulations also show that birth radius can be predicted with a simple regression model using predictors of ([Fe/H],age) alone (see \citealp[][]{Ratcliffe2020}) and that in the [Fe/H]-[Mg/Fe] plane, birth radius and age are ordered in their distribution across this plane (see Figure 3 in \citealp[][]{Buck2020}). This assumption motivates the way we have dissected the data (i.e., Figures 4 -7) and why we have selected to employ representative stars, in bins of ([Fe/H],[Mg/Fe], age), in order to undertake our chemical evolution modelling procedure. 

We note that we do not seek to lay out any analytical formalism for the ([Fe/H],[Mg/Fe],age) relationship with birth radius. Instead, we use this philosophy to guide our modelling approach. In the following sections, we will go through our method for determining representative stars that will later model utilising the chemical cells (or MAPS) method.

\subsection{Empirical Mapping of Neutron Capture Trends} 

We adopt a novel approach to study and interpret the age-neutron capture element abundance relations in the Milky Way disc using a grid of chemical cells in the [Fe/H]-[Mg/Fe] plane. In detail our procedure is as follows :\\
\\
i) We dissect our parent sample of stars into 5x5 grid of chemical cells in the [Fe/H]-[Mg/Fe] plane (where Mg is our representative $\alpha$-element) in order to study the abundance-age profiles of Galactic disc populations with similar SNII-to-SNIa ratio contributions. While we select a 5x5 grid across the [Fe/H]-[Mg/Fe] plane, we find that the number of cells could be increased/decreased to select more/fewer stellar populations. However, we validate that employing a 3$\times$3 and/or 10$\times$10 grid, finding that our results remain unchanged. To quantify the precision of our chemical abundance results we run a sampling of the data given the errors, assuming Gaussian distributed distributions:
\begin{itemize}
    \item For every star observed in our parent sample, we draw 200 samples of their respective [Mg/Fe] and [Fe/H] abundances from a normal distribution that is centred at the observed value, and whose standard deviation equates to the uncertainty in the abundance measurement (provided by the GALAH catalogue). This sampling procedure yields the equivalent of 200 [Fe/H]-[Mg/Fe] chemical composition planes for every star observed, and permits one to assess the impact of the measurement uncertainties in the abundances, and to account for stars moving between different cells in our fixed grid. The resampling is, in effect, a test of the stability of our results given our bin edges. By resampling the points in their errors, we can test the sensitivity to the uncertainties of the points lying near the bin boundaries. We utilised N=200 here as a large enough number do this test thoroughly (although far fewer would have revealed if our results were sensitive to our bin boundaries given the data). The typical standard deviation on the mean value of ([Fe/H],[Mg/Fe]) in a bin from 200 samples is (0.0008,0.0007) dex. Therefore, our results are clearly stable and insensitive to the bin edges. Nevertheless, we include this for completeness.
    \item For each of the 200 samples of every star (thus each of the 200 [Fe/H]-[Mg/Fe] realisations of the chemical composition plane), we dissect the chemical plane into a 5$\times$5 grid and loop over every chemical cell. This grid covers a range in metallicity from --1 < [Fe/H] < 0.5, and a range in [$\alpha$/Fe] of --0.4 < [Mg/Fe] < 0.6. For every iteration in the loop, we only select those stars that reside within each individual cell as our individual chemical cell samples. 
    \item Upon selecting the stars in each chemical cell (for every 200 sample iterations of the [Fe/H]-[Mg/Fe] plane), we divide the populations by stellar age into seven equally spaced bins, ranging from 0 to 14 Gyr (each covering a range of 2 Gyr). The mean age uncertainty is on the order of ~2-3Gyr. Therefore, for every sample and every chemical cell, we have seven samples of stars binned by age. The 200 draws of the data given the measurement uncertainties per cell are used to determine the mean values along the age-abundance relations and the confidence around these trends. These ([Fe/H], [Mg/Fe], age) bins are also utilised to report the mean orbital properties across the age-abundance relations and to select the mean abundance values for three age samples per cell, that we will use later for our chemical modelling procedure.
\end{itemize}

ii) We subsequently examine the age-abundance relations for [Ba/Fe] and [Eu/Fe] in each cell; [Ba/Fe] is our $s$-process tracer element, and [Eu/Fe] is our $r$-process tracer. In each cell, we fit a running mean through the age-[Ba/Fe] and age-[Eu/Fe] data. We use the mean and $1-\sigma$ standard deviation around the mean, for each of our 200 samples, to report the overall mean age-abundance trends and the 1-$\sigma$ dispersion around these trends (and return the corresponding confidence intervals). We note that the standard error on the mean age-abundance trends is negligible and is smaller than the width of the mean trend. We also determine the slope of the relation for the running mean for stars with an age between 2 and 10 Gyr, and calculate its associated uncertainty by determining the slope for every one of the 200 samples and determining the standard deviation.\\

iii) We explore how the chemical signatures of stellar populations binned in ([Fe/H], [Mg/Fe], age) space connect to the spatial distribution of these stellar populations by examining the relationship between the age-[Ba/Fe] gradients and the mean orbital radius of stars across chemical cells (Section~\ref{results_radius}).

\subsection{Modelling the data with Chempy: a flexible Galactic chemical evolution model}
\label{chempy_definition}

We model the observational results using flexible GCE models (namely, Chempy: \citealp[][]{Rybizki2017}), and use our empirical findings to constrain our theoretical understanding of the origin source of neutron capture elements and the chemical evolution of the Galactic disc. Broadly speaking, Chempy \citep[][]{Rybizki2017} is a flexible Galactic chemical evolution (GCE) model that serves as a link between the parameters of a chemical evolution model $\theta$ (e.g., the IMF or SNIa explosion rate) together with underlying hyperparameters $\lambda$ (e.g., a specific yield set) to the likelihood of observations $\mathcal{O}$ (such as stellar abundances) via its model predictions $\delta$. Therefore, the goal of Chempy is to be able to predict observed abundance patterns of stars throughout cosmic time by inferring, via Bayesian inference, key GCE model parameters by employing physically motivated models of star formation and stellar evolution together with a parameterization of galactic ISM physics. 

In more detail, Chempy is a one-zone GCE model with seven free parameters. Three of these are linked to stellar physics, and include the high-mass slope of the IMF ($\alpha_{\mathrm{IMF}}$), a normalisation constant for the number of SNIa explosions (log$_{10}$(SNIa)), and a parameter for the time delay of SNIa enrichment (log$_{10}$($\tau_{\mathrm{I_{a}}}$)). The remaining four parameters are related to the interstellar medium (ISM), and include a parameter that governs the star formation efficiency (log$_{10}$(SFE), defined as SFE = $m_{\mathrm{SFR}}$/$m^{n_{\mathrm{Schmidt}}=1}_{\mathrm{ISM}}$), where $m_{\mathrm{SFR}}$ is the mass formed from a star formation rate (within a given time), and $m^{n_{\mathrm{Schmidt}}=1}_{\mathrm{ISM}}$ is the mass of the ISM assuming a linear Schmidt law $n_{Schmidt}=1$ infall of gas needed to sustain a certain star formation rate; but see also the definition given in \citet{Prantzos2018}), a parameter that controls the peak of the star formation rate (SFR$_{\mathrm{peak}}$), the fraction of stellar yields which outflow to the surrounding gas reservoir ($\chi_{\mathrm{out}}$), and the initial mass of the gas reservoir (log$_{10}$($f_{\mathrm{corona}}$)). With these parameters controlling the GCE model, Chempy calculates the enrichment of the ISM through cosmic time utilising a set of yield tables that describe the enrichment from three nucleosynthetic channels: SNII, SNIa, and AGB stars. Of these, the SNII and AGB yields are mass and metallicity dependent, and in Chempy these yields are ejected immediately following stellar death. Conversely, SNIa enrichment is mass and metallicity independent, so yields are deposited into the ISM according to the \citet[][]{Maoz2010} delay time distribution. 

For an in-depth description of the intricacies of Chempy, a complete description of its model parameters, and the assumptions made in the models, we refer the reader to \citet[][]{Rybizki2017}.

One of the main advantages of Chempy is the ability to constrain the parameters from the GCE model using a set of observational measurements. This is achieved by inferring the posterior distribution of the GCE model parameters in a defined parameter space, given the data, using MCMC sampling (using the \texttt{emcee} package \citep[][]{Foreman2013}).
 The MCMC implementation enables an initialization of a collection of walkers for every parameter fitted in the model, using an evaluated logarithmic prior value for the current parameter configuration. Each walker then evaluates the posterior at every step of its random walk, and compares the current evaluation of the logarithmic likelihood with that of its previous step in order to assess if the step has helped improve the fit to the data, where the proposed step is then either accepted or rejected with some probability. Then, keeping a record of the chemical abundances at each timestep, the Chempy GCE is run with these parameter values from $t$=0 to the provided estimated time of stellar birth of the star being modeled. The log-likelihood is calculated using a $\chi^{2}$ procedure by comparing the predicted abundances to that of the model given the best fit parameters, and the posterior is computed by adding the log-prior to the log-likelihood.
 
For this work, we use Chempy to individually fit the measurements of ([Fe/H], [Mg/Fe], [Ba/Fe], age) for twenty-seven representative stars. To define these representative stars, we first construct a grid of nine cells in the [Fe/H]-[Mg/Fe] plane, as shown in Figure~\ref{mgfe}. The [Fe/H]-[Mg/Fe] cells  contain the majority of our disc sample, and span a range in [Mg/Fe] of --0.2 < [Mg/Fe] < 0, 0 < [Mg/Fe] < 0.2, and 0.2 < [Mg/Fe] < 0.4, and a metallicity range of --0.7 < [Fe/H] < --0.4, --0.4 < [Fe/H] < --0.1, and --0.1 < [Fe/H] < 0.2. The [Fe/H] and [Mg/Fe] measurements of the representative stars are the means of each bin. We further break this up into three age ranges, and take the mean of each bin as the age of each representative star. These three bins comprise an old (i.e., 8 < Age < 10 Gyr), an intermediate (i.e., 5 < Age < 7 Gyr), and a young (i.e., 2 < Age < 4 Gyr). We calculate the mean [Ba/Fe] in each bin of [Fe/H], [Mg/Fe] and age.  We could have instead selected individual stars, or made bins in four dimensions (so including [Ba/Fe]). However, with our ordered approach, we (i) systematically sample the extent of the environment (ii) leverage the large sample of data and isolate regions of similar birth environment using [Fe/H], [Mg/Fe] and age alone. For a typical representative star there are between $\sim$50 to $\sim$5000 real stars that fall within the boundary limits. Within each chemical cell, there are from $\sim$200 to $\sim$14,000 stars.

We highlight that because we are fitting individual Chempy models for each representative star in ([Fe/H],[Mg/Fe],[Ba/Fe],age) space, we are closer to the one-zone model concept. This is because our grid of cells in ([Fe/H],[Mg/Fe],age) is intended to each isolate individual birth environments. The stellar populations in these cells can be presumed to have been born in similar regions of the Galaxy (\citealp[e.g.,][]{Frankel2018,Minchev2018,Ness2019,Ness2021}). Therefore, our approach is designed not to mix populations, but to isolate populations of common birth time and radius to input to the chemical evolution model.

To summarise, as inputs into the Chempy model we use mean values of [Mg/Fe], [Fe/H], [Ba/Fe], and age, as the representative values of the corresponding cell, determined using the sampling technique described above. For the uncertainties on these values, we utilise the mean value of the measurement errors (namely, [Mg/Fe]$_{\mathrm{err}}$, [Fe/H]$_{\mathrm{err}}$, [Ba/Fe]$_{\mathrm{err}}$, and Age$_{\mathrm{err}}$). The ability to fit the chemical abundance data in small bins of ages and chemical space brings our methodology closer to the one-zone GCE model concept. Using these measurements, Chempy optimises for four parameters of the environment; 
type Ia progenitors, the Star Formation Rate (SFR), the Star Formation Efficiency (SFE) and the high-mass slope of the Initial Mass Function (IMF).

Each run in Chempy is performed assuming the same initial conditions for the ISM (see Fig 1 from \citealp[][]{Rybizki2017} for details), assuming a $\tau_{\mathrm{start}}$=13.5 Gyr. Thus, Chempy attempts to match the observed abundance data by running the GCE for a time of $\tau$ = $\tau_{\mathrm{start}}$-age. It does this by searching the parameter space to get the most likely set of model parameters that fit the data, quantified by minimising a $\chi^{2}$ value. For each run of the Bayesian inference of the model parameters, we adopt the same parameter priors, limits, and number of walkers/steps.
A summary of these values can be found in Table~\ref{tab_chempy}. In brief, we fit every stellar population by initialising a MCMC routine with 20 walkers, and run the MCMC procedure for 1,000 steps. At the beginning of the routine, the walkers start from the default Chempy prior values, and explore a wide range of parameter space before converging and yielding the resulting optimized model parameters. We find that on average, each stellar population fitted converges to the resulting optimized parameter values after approximately 300 steps in the chain. An example of the resulting MCMC samples for all three age bins within the chemical cell spanning --0.1 < [Fe/H] < 0.2 and --0.2 < [Mg/Fe] < 0 is shown in Appendix~\ref{app_mcmc}.

\section{Results}
\label{results}

\subsection{Mapping neutron-capture abundances in a supernovae reference plane}
\label{sec_cells}

The [Fe/H]-[Mg/Fe] is a familiar reference frame for the exploration of additional abundances. The grid across this plane comprises our chemical cells: this grid spans a range in [Fe/H] of --1 < [Fe/H] < 0.5, increasing in steps of 0.3 dex, with a range in [Mg/Fe] from --0.4 < [Mg/Fe] < 0.6, increasing in 0.2 dex steps. Across the [Fe/H]-[Mg/Fe] plane, an individual chemical cell represents a group of stars with the same total ratio of enrichment by SNII and SNIa. The ability to isolate stellar populations that likely formed under similar conditions in this manner enables the opportunity to study properties (such as abundance-age relations and orbits) that are independent from the main source production sites of the elements comprising the chemical abundance plane grid. For the case of the [Fe/H]-[Mg/Fe] plane, it provides the opportunity to study properties that do noot depend on the amount of enrichment due to SNII and SNIa.

\begin{figure*}
    \centering
    \includegraphics[width=\textwidth]{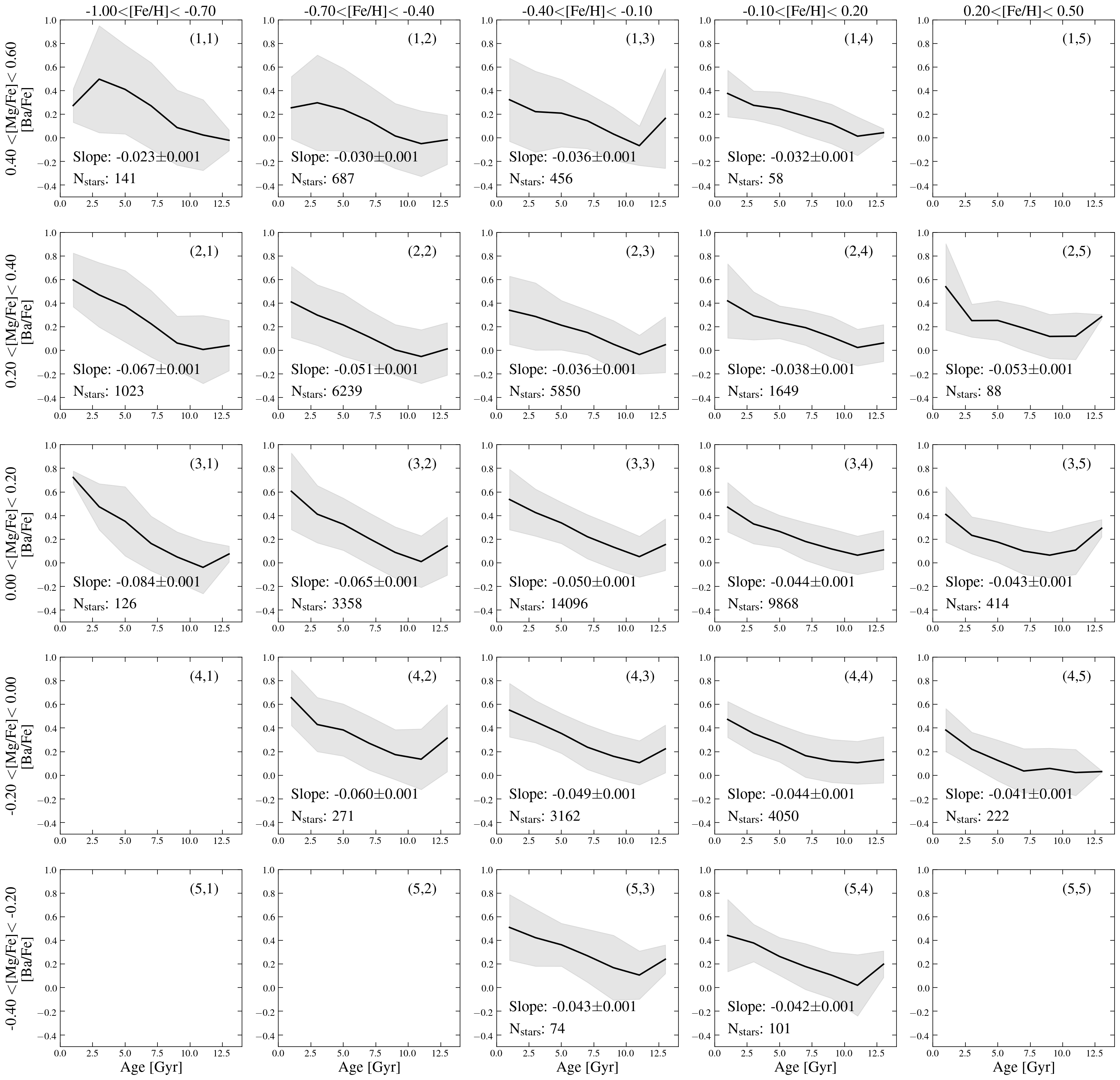}
    \caption{Age-[Ba/Fe] relations for stars in the chemical cells gridded in the [Fe/H]-[Mg/Fe] plane. Here solid line is the running mean value, and shaded regions correspond to the 1-$\sigma$ dispersion around the line. Abundance-age profiles for cells containing fewer than 20 stars are removed. Also shown in each cell is the number of stars contained and the slope of the abundance-age relation between 2 and 10 Gyr. Across all cells, stellar populations at fixed ([Mg/Fe],[Fe/H]) that are younger present higher [Ba/Fe] values than their older counterparts. The numbers in the top right corner correspong to a matrix index for each cell in the grid. We note that even when excluding the (contentious) young high-alpha stars from cell (1,1), and the old low alpha stars from cells (4,5) and (5,4), our results survive.}
    \label{bafe_cells}
\end{figure*}

\begin{figure*}
    \centering
    \includegraphics[width=\textwidth]{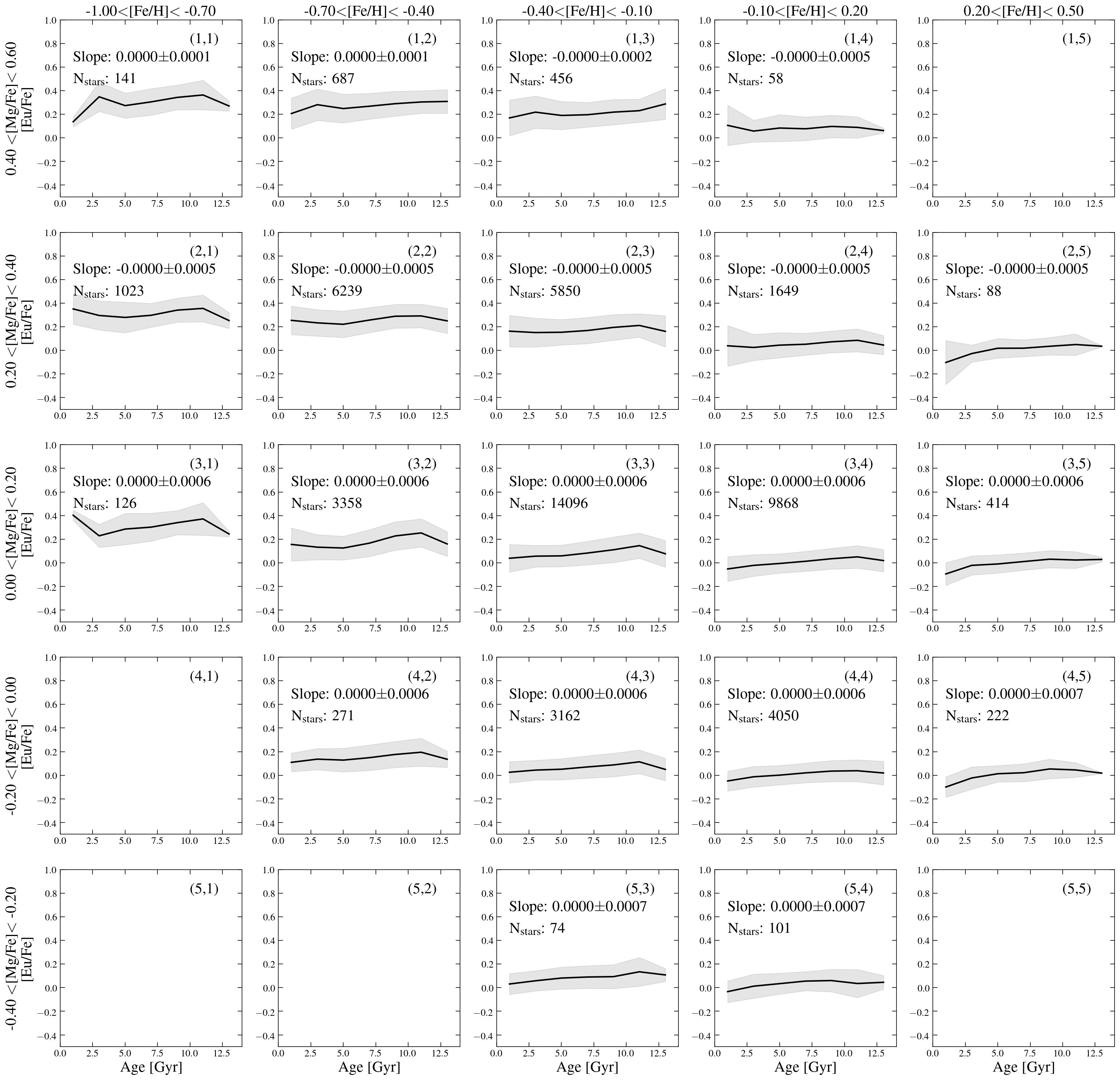}
    \caption{Same as Fig~\ref{bafe_cells} but now for the age-[Eu/Fe] relations. In contrast to the results from Fig~\ref{bafe_cells}, the age-[Eu/Fe] profiles are approximately flat for the majority of the chemical cells (with the exception of the lowest [Fe/H] and [Mg/Fe] cells, likely due to contamination from accreted populations at the oldest age bins).}
    \label{eufe_cells}
\end{figure*}

The mean age-[Ba/Fe] and age-[Eu/Fe] profiles in the chemical cell plane are shown in  Fig~\ref{bafe_cells} and Fig~\ref{eufe_cells},
respectively, where only cells containing more than 20 stars are shown. The age-abundance trends are shown as a running mean of the data, and the shaded band is the confidence on this mean (calculated as the 1-$\sigma$ dispersion). In addition, we also show the slope value obtained when fitting the running mean with linear regression for stars between 2 and 10 Gyr of age and its associated uncertainty, calculated using the running mean (abundance,age) values and by calculating the mean of the standard deviation of the slope computed in the 200 iterations, respectively. We ensured our age-abundance relationships were not affected by any unforeseen systematic effects due to trend in stellar parameters by checking the abundance-age profiles in a very narrow T$_{\mathrm{eff}}$, and found that the results remained consistent.

Connecting the assumption of ([Fe/H], [Mg/Fe], age) being a link to birth radius back to Figures \ref{bafe_cells} and \ref{eufe_cells}, this means that moving along age within a cell, as well as at fixed age and moving between cells, is equivalent to looking at populations formed at a different birth radius in the Milky Way. We make no quantitative assumptions here about the precise form of this relation (see \citet[][]{Frankel2018} and \citet[][]{Minchev2018} for this), as our goal is not to infer birth radius quantitatively. Rather, we want to isolate groups of stars that were presumably born together in order to undertake an empirical assessment of what the neutron-capture abundances are for stars that are born at the same radius (with some presumably small dispersion, i.e. at fixed ([Fe/H],[Mg/Fe],age). This is what is captured in examining the age-[Ba/Fe] and age-[Eu/Fe] trends in small bins of ([Fe/H],[Mg/Fe]).

\begin{figure*}
    \centering
    \includegraphics[width=\textwidth]{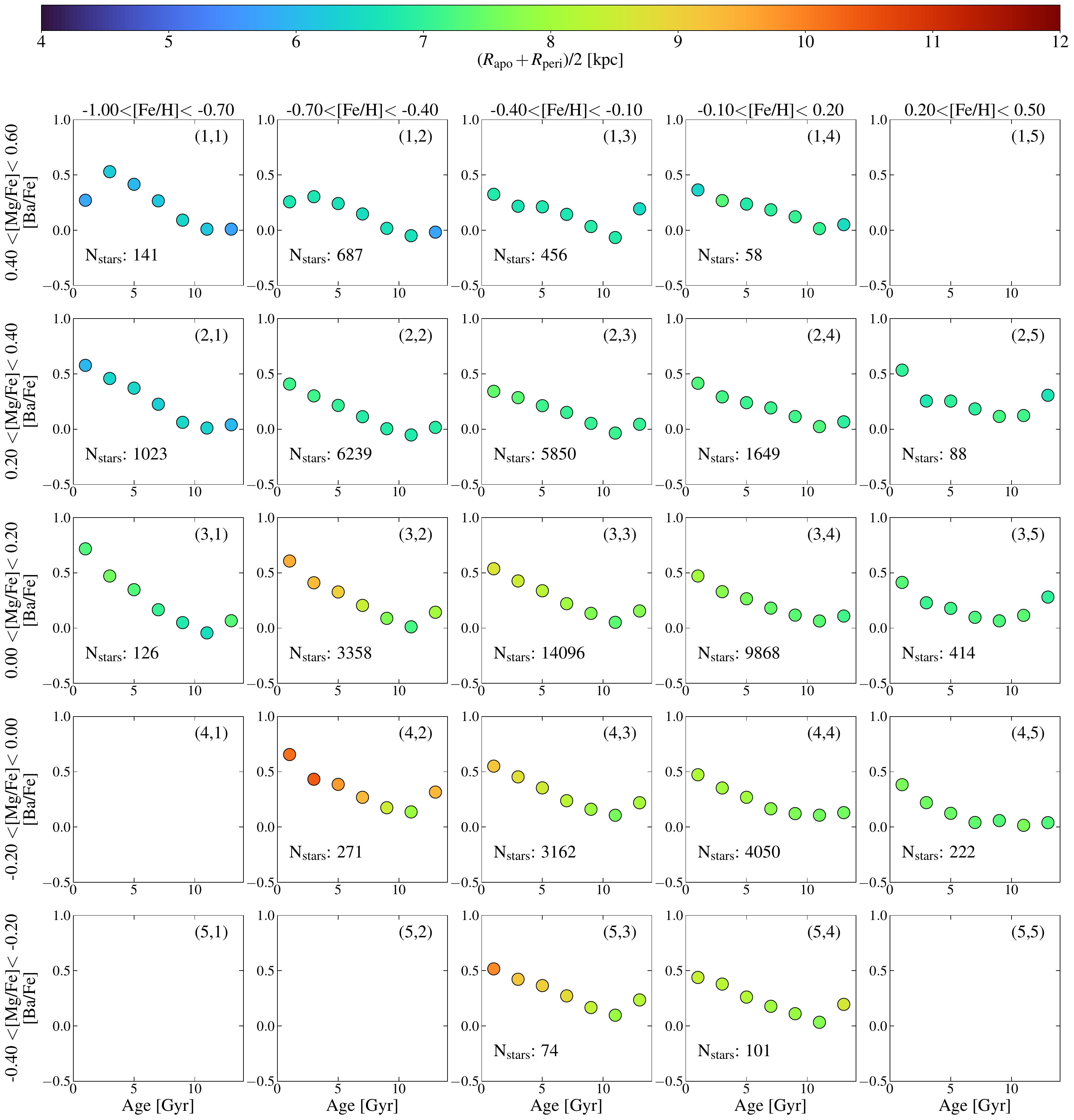}
    \caption{Same as Fig~\ref{bafe_cells}, but now colour coded by mean orbital radius. It is evident that the mean orbital radius of a stellar population directly affects the number of contributions a stellar population receives from the winds of low/intermediate-mass AGB stars (namely, the amount of Ba). }
    \label{bafe_cells_rmean}
\end{figure*}

\subsubsection{The $s$-process element: Barium}
\label{sec_bafe}

Focusing on the mean age-[Ba/Fe] relations that are shown in Fig~\ref{bafe_cells}, we see clear [Ba/Fe]-age trends in all [Fe/H]-[Mg/Fe] cells across the disc of the Galaxy. Here, only cells containing more than 20 stars are illustrated. For every chemical cell displaying a [Ba/Fe]-age relation in the grid, we observe stars across all age bins (namely, from an age of $\sim$1 Gyr to $\sim$13 Gyr) and therefore report an age-abundance relation in every one of these cells. We report that the young high-$\alpha$ and [Fe/H]-poor stars identified and studied previously in the Galaxy exist in our sample (\citealp[e.g.,][]{Chiappini2015,Silva2018,Hekker2019,Miglio2021,Zinn2021, Zhang2021}). \citet[][]{Chiappini2015} have postulated that these are formed close to the Galactic Centre, likely close to the end of the Galactic bar.  We find these stars in our sample to have a mean orbital radius between 4-6 kpc (see Fig~\ref{bafe_cells_rmean}). However, we note that recent work argues that these young high-$\alpha$ stars are likely intrinsically old and have anomalous surface abundances and properties that make them only appear young \citep[][]{Zhang2021}.

In a similar fashion, we also observe old low-[Mg/Fe] stars in our sample. We note that these stars are similarly found in the APOGEE survey (see \citealp[][]{Bovy2019,Hasselquist2019,Sit2020}) as well as in the high-fidelity Kepler asteroseismic sample (\citealp[e.g.,][]{Silva2018,Miglio2021}). These stars appear to originate from the inner Galaxy (see the [Fe/H]-[$\alpha$/Fe] plane in \citealp{Hayden2015} and \citealp[][]{Eilers2021}), and are expected to have moved into the present day bin via radial migration, which is measured to be significant in the Milky Way (\citealp[][]{Frankel2018,Minchev2018}) and predicted from cosmological simulations \citep[][]{Yuxi2021}. We note that if we were to exclude these young high-$\alpha$ stars, as well as the old low-$\alpha$ stars, our results would remain unchanged.

Within each cell studied, the [Ba/Fe] decreases with increasing age: old stars show lower [Ba/Fe] abundances than young stars. The amplitude of the slope of the relation is consistent with previous studies \citep{Nissen2015, Bedell2018, Sales2021, Zinn2021}, and has a mean value of $\sim$--0.05 dex/Gyr across all cells for stellar populations with an age between 2 and 10 Gyr.

Across cells, the slope of the age-[Ba/Fe] relations changes in a smooth and ordered way. When comparing the [Ba/Fe] relations between chemical cells at fixed [Fe/H], we see that the abundance-age relations undergo a slight "shift" to higher [Ba/Fe] values across all ages with decreasing [Mg/Fe]. This can be clearly observed in the --0.4 < [Fe/H] < --0.1 column of cells, where we find that the age-[Ba/Fe] relation for stellar populations in the 0.4 < [Mg/Fe] < 0.6 is lower than the relation displayed in the --0.4 < [Mg/Fe] < --0.2 counterpart cell, shifted down by approximately 0.2 dex across all ages. We also find that on average, for fixed [Mg/Fe], the age-[Ba/Fe] relation slope becomes slightly flatter with increasing [Fe/H]. Here, the old populations across cells display a similar Ba value of [Ba/Fe]$\sim$0.15 for an age of $\sim$10 Gyr. However, the young populations show a sharper drop in [Ba/Fe] with increasing [Fe/H]. This is clearly seen in the middle row (namely, at 0 < [Mg/Fe] < 0.2), where the [Ba/Fe] abundances of stellar populations with an age of $\sim$ 2.5 Gyr drop from a value of [Ba/Fe]$\sim$0.6 at --1 < [Fe/H] <
--0.7 to a value of [Ba/Fe]$\sim$0.2 at 0.2 < [Fe/H] < 0.5. 

In addition to the clear differences between the age-[Ba/Fe] relations (both within each individual cell and across cells), there is a clear upturn of [Ba/Fe] at the oldest ages in most of the chemical cells. This feature implies an an elevated amount of Ba almost at the beginning of the Milky Way's chemical evolution history, which is unexpected given that Ba is an $s$-process element primarily synthesized by AGB stars that have a late onset time. We note that this upturn feature at the oldest ages has also been reported in an analysis of GALAH DR3 red giant data by \citet[][]{Hayden2020, Zinn2021}. \citet[][]{Hayden2020} suggest that this upturn at the oldest ages is due to difficulties in determining $s$-process elements for very old metal-poor stars in the high-$\alpha$ disc. It is possible that the upturn in [Ba/Fe] at the oldest ages could be due to underlying effects in the determination of Ba for very old red giant stars. However, we find that this upturn in [Ba/Fe] at old ages is present in almost all the chemical cells studied, not just in the metal-poor and [$\alpha$/Fe] enhanced cells. Given this striking result, we hypothesise another possible explanation to explain this sharp feature. We speculate that the clear upturn in [Ba/Fe] at old ages in Fig~\ref{bafe_cells} could be due to the contribution of AGB winds (namely, the main nucleosynthetic process of Ba) having a more delayed onset when compared to SNIa and SNII. In this scenario, the primordial [Ba/Fe] at the oldest ages is decreased by the Fe contribution of SNIa, which has a slightly earlier onset time than the contribution from AGB winds. This would lead to a decrease in the [Ba/Fe] value at the oldest ages within a window of [Fe/H], as observed in Fig~\ref{bafe_cells}. The decrease in [Ba/Fe] would then last for the time frame between the onset of SNIa and AGB winds. After this, Ba begins to be synthesised in the winds of low/intermediate AGB stars, leading to an increase in [Ba/Fe] towards younger ages as observed in our [Ba/Fe]-age profiles. While this scenario is speculative, it is possible to reconcile the trends observed in Fig~\ref{bafe_cells} under this hypothesis. Moreover, recent theoretical results predict such a behaviour in the evolution of [Ba/Fe] at the oldest ages \citep[][]{Kobayashi2020b}, reinforcing the notion that this feature is likely physical. Furthermore, we also speculate that the reason for this abrupt upturn in [Ba/Fe] at the oldest ages across all cells may indicate that the oldest disc stars (Age $>$ 10 Gyr) may not have formed with the same radius-[Fe/H] gradients as the rest of the disc \citep[][]{Sales2021}. 

In any case, we note that such features do not affect our modelling procedure when fitting the data with the Chempy GCE model (see Section~\ref{sec_modelling}) as we choose to only utilise [Ba/Fe] and age measurements for stars with ages between 2-10 Gyr in our analysis. Thus, we suggest that this unexpected upturn at the oldest ages should be further investigated in future studies. 

The age-[Ba/Fe] gradients in Figure~\ref{bafe_cells} indicate that the Ba nucleosynthetic sources yield different amounts of [Ba/Fe] at different times, at fixed SN contributions (which provide the majority of Mg and Fe abundances). This is not unexpected, as Ba is primarily synthesised via the accretion of stellar winds from low-/intermediate-mass asymptotic giant branch (AGB) stars (\citealp[e.g.,][]{Sneden2008,Karakas2016, Escorza2020, Kobayashi2020b}). Therefore, the genesis of this element is primarily independent of SN contributions.

In Appendix~\ref{app_scatters} we also show the scatter around the mean age-abundance relations $\sigma_{\mathrm{[Ba/Fe]}}$-age relations for every [Fe/H]-[Mg/Fe] chemical cell in Fig~\ref{bafe_cells_uncer}. Our results show that the mean $\sigma_{\mathrm{[Ba/Fe]}}$-age relations are predominantly flat across all chemical cells. The amplitude of this scatter and its (lack of) variance over age is presumably a constraint on to the chemical evolution history of the Galaxy disc: whatever sets the amplitude of the slope as well as the scatter around it is fairly insensitive to stellar age. 

\subsubsection{The $r$-process element: Europium}
\label{sec_eufe}

Following our detailed inspection of the age-[Ba/Fe] profiles, we now draw the reader's attention to Fig~\ref{eufe_cells}, where we show the [Eu/Fe]-age relations for disc populations in the same [Fe/H]-[Mg/Fe] chemical cells grid from Fig~\ref{bafe_cells}. The [Eu/Fe]-age relations show an almost flat profile across all chemical cells, that can be quantified given the small slope value determined across all cells studied (namely, of $\sim$0.004). Note there are some small variations for old ages at the lowest [Fe/H] (and [Mg/Fe]) cells, possibly caused by the contamination of the most metal-rich populations belonging to the accreted debris from the \textit{Gaia}-Enceladus/Sausage accretion event \citep[][]{Belokurov2018,Haywood2018,Helmi2018,Mackereth2019}, which has been recently shown to host stellar populations with enhanced $r$-process elements (\citealp[e.g.,][]{Aguado2021,Matsuno2021}), or other accreted populations recently discovered (\citealp[e.g.,][]{Myeong2019,Horta2021}). The completely different profile of [Eu/Fe] compared to [Ba/Fe] is indicative  that these two elements are synthesised in different ways. This result is somewhat not unexpected, as Eu is an $r$-process neutron capture element, whereas Ba is an $s$-process element. The flat [Eu/Fe]-age relations across all chemical cells from Fig~\ref{eufe_cells} suggests that Eu must be synthesised similarly to Mg and/or Fe via the detonation of SNII/SNIa, for the disc population. This result corroborates the hypothesised origin source of Eu in exploding massive stars \citep[][]{Kobayashi2020b}.

Figure \ref{eufe_cells} shows that while the age-[Eu/Fe] relations are flat, they vary in amplitude at fixed [Mg/Fe], across [Fe/H]. The overall [Eu/Fe] abundance shifts to lower values across all ages in a cell with increasing [Fe/H]. This can be clearly seen by focusing on the 0.2 < [Mg/Fe] < 0.4 row, where we find that the [Eu/Fe] values across all ages in the --1 < [Fe/H] < --0.7 cell have [Eu/Fe]$\sim$0.3. This is approximately $\sim$0.3 dex  higher than the [Eu/Fe]$\sim$0 value in the 0.2 < [Fe/H] < 0.5 cell, despite these displaying an almost identical flat [Eu/Fe]-age profile. We postulate that the reason for this observed behaviour is due to Eu having no strong dependence on SNIa contributions. This would imply that with increasing [Fe/H] the ratio of [Eu/Fe] gets diluted (although at a different rate to which [Fe/H] increases, as there will still be Eu being produced, in different amounts), making the overall [Eu/Fe] abundance value drop equally across all ages, as is observed in Fig~\ref{eufe_cells}. Conversely, we find that [Eu/Fe] varies --at fixed [Fe/H]-- with [Mg/Fe]. For example, at --0.7 < [Fe/H] < --0.4, the mean value of [Eu/Fe] drops from $\sim$0.3 at 0.2 < [Mg/Fe] < 0.4 to $\sim$0.1 for 0 < [Mg/Fe] < 0.2. We speculate this is likely due to the exploding massive star origin source, from which Mg is also primarily produced.

In an identical fashion to Fig~\ref{bafe_cells_uncer}, we show the scatter in the age-[Eu/Fe] relations in Fig~\ref{eufe_cells_uncer} in Appendix~\ref{app_scatters}.

\subsection{Orbital properties across the grid of age-[Ba/Fe] relations}
\label{results_radius}
We have examined the [Ba/Fe] and [Eu/Fe] abundances with age across the [Fe/H]-[Mg/Fe] plane. To explore the empirical relations between the chemical to the structural architecture of the disc we examine the [Ba/Fe] relations, for which we see gradients with age within cells, as a function of mean orbital radius. Fig~\ref{bafe_cells_rmean} shows the same age-[Ba/Fe] abundance profiles for the [Fe/H]-[Mg/Fe] chemical cells presented in Fig~\ref{bafe_cells}, but now colour coded by their mean orbital radius value. This is defined as the sum of the pericentric and apocentric distances divided by two, namely, ($R_{\mathrm{apo}}$ + $R_{\mathrm{peri}})$/2. It is  evident from Fig~\ref{bafe_cells_rmean} that there are remarkable gradients in the mean orbital radius across the [Fe/H]-[Mg/Fe] plane \citep[i.e. also see][]{Hayden2015}, and, within some cells, along the age-[Ba/Fe] relations. The structure that we see here is an integration of the birth location of the stars and their migration (and vertical heating) over time (\citealp[e.g.,][]{Ness2019,Buck2020,Yuxi2021}). We find looking across the plane, that stellar populations of a given age and [Fe/H] decrease in mean orbital radius with increasing [Mg/Fe]. For chemical cells with [Mg/Fe] $>$ 0.2, [Fe/H] and [Mg/Fe] alone are indicative of the present day guiding radius. However, in some cells, with [Mg/Fe] $<$ 0.2 and -0.7 $>$ [Fe/H] $>$0.2, the Ba abundance and age are correlated with the mean guiding radius; younger and more [Ba/Fe]-rich stellar populations typically have larger orbital radius values than their older counterparts. See Fig~\ref{bafe_rmean_age} in the Appendix~\ref{app_bafe_rmean} for further details.

\section{Chemical Evolution modelling of the age-[Ba/Fe] relations across the disc} 
\label{sec_modelling}

So far, we have studied the empirical age-[Ba/Fe] and age-[Eu/Fe] relations of Galactic disc populations in different [Fe/H]-[Mg/Fe] chemical cells. Empirically, we see in Figures \ref{bafe_cells} and \ref{eufe_cells}, that across the chemical grid of ([Fe/H],[Mg/Fe]), when we examine the age-[Ba/Fe] and age-[Eu/Fe] relations, the slope and/or amplitude of these relations changes with both [Fe/H] and [Mg/Fe]. From the mono-abundance plane of \citet[][]{Bovy2012} showing the different scale lengths and heights for stars in the Galactic disc as a function of their ([Fe/H]-[Mg/Fe]) cell, as well as our own work which shows the age-abundance relations changing at different locations in [Fe/H], [Mg/Fe] (see also \citet[][]{Ness2019} and \citet[][]{Yuxi2021}) we believe that it is a reasonable and sound assumption that a cell (or small bin) in ([Fe/H], [Mg/Fe], age) isolates a population of stars that likely were born together at a particular radius and time in the Milky Way disc. In more detail, we see from Figures \ref{bafe_cells} and \ref{eufe_cells} that the age-[Ba/Fe] relations flatten in slope with increasing [Fe/H] across chemical cells of fixed [Mg/Fe], and increase in amplitude with [Mg/Fe] across cells with fixed [Fe/H]. Conversely, abundance gradients across age within chemical cells are not observed in the [Eu/Fe]-age profiles. Moreover, in Fig~\ref{bafe_cells_rmean}, we have seen that for some of the chemical cells, the [Ba/Fe]-age profiles also have a noticeable relation with orbital radius. In cells with [Mg/Fe] $<$ 0.2, the younger populations of a given chemical cell typically have higher mean orbital radii values when compared to their older counterparts (see also Fig~\ref{bafe_rmean_age} in Appendix~\ref{app_bafe_rmean}). For example, in the cell containing stellar populations covering 0 $<$ [Mg/Fe] $<$ 0.2 and $-0.7 < [Fe/H] < -0.4$, we see that the oldest populations show a [Ba/Fe] $\sim$ 0 and an orbital radius of ($R_{\mathrm{apo}} + R_{\mathrm{peri}}$)/2 $\sim$ 7 kpc. The youngest populations show a [Ba/Fe] $\sim$ 0.65 and a ($R_{\mathrm{apo}} + R_{\mathrm{peri}}$)/2 $\sim$ 9.5 kpc. Conversely, for chemical cells with [Mg/Fe] > 0.2, we find that these orbital radius-[Ba/Fe] relations are flat. 

 In light of all these results, and under the assumption that Ba is primarily released in the winds of low-/intermediate-mass AGB stars, we reason that the variation in the [Ba/Fe] abundance with age within and between the [Fe/H]-[Mg/Fe] chemical cells is likely due to a difference in the total number of low-/intermediate-mass AGB stars, that contribute $s$-process (Ba) to the ISM. This enrichment in Ba is a function of stellar age, but is also a function of birth radius across the disc, given that we see the [Ba/Fe]-age relations change for different [Fe/H]-[Mg/Fe] cells. Stars of different ages have either different formation pathways or form in different environments at a different time from different gas. This enables the same ([Fe/H],[Mg/Fe]) but different [Ba/Fe] values to be reached.
 
 As highlighted by \citet[][]{Ting2021}, there is a marginal amount of additional information in individual abundances [X/Fe] at fixed ([Fe/H],[Mg/Fe]). The marginal additional information provided by additional chemical abundances potentially are a key marker or discriminator of the environment in which a stellar population in born in the Galaxy. Here, we access this marginal but key information in using the corresponding [Ba/Fe] abundance at fixed ([Fe/H],[Mg/Fe],age) to examine what GCE model parameters are consistent with each of the different points in the chemical-age hyper-plane. As we are utilising [Fe/H], [Mg/Fe], and [Ba/Fe] as our input abundances (as well as age) for stellar populations contained within different [Fe/H]-[Mg/Fe] cells, we are able to tap into the nucleosynthetic sources of SNIa, SNII, and AGB across different birth times and radii in the Galaxy. Thus, we analyse the full range of the ([Fe/H],[Mg/Fe],age) plane of the Milky Way disk.

To interpret the empirical data we use the flexible evolution code Chempy \citep[][]{Rybizki2017} (see Section~\ref{chempy_definition} for further details). In brief, Chempy is a Bayesian approach to infer the posterior distribution of the GCE model parameters in a defined space given the data using a MCMC procedure. Using Chempy, we determine if we can fit the observed joint [Ba/Fe], [Fe/H] and [Mg/Fe] element abundances, as well as age, by allowing the GCE model parameters to vary. We note that as the $r$-process nucleosynthesis channel is not included in the current Chempy models, we are unable to model the [Eu/Fe]-age profiles. Of the set of Chempy's GCE model parameters, we explore and optimise the following: the high-mass slope of the IMF ($\alpha_{\mathrm{IMF}}$), the normalisation constant for the number of SNIa explosions (log$_{10}$(SNIa)), the star formation efficiency (log$_{10}$(SFE)), and the peak in the star formation rate (log$_{10}$(SFR$_{\mathrm{peak}}$)). The parameters we chose to exclude from the optimisation routine are the time delay of SNIa enrichment (log$_{10}$($\tau$I$_{\mathrm{a}}$)), the fraction of stellar yields which outflow to the surrounding gas reservoir ($\chi_{\mathrm{out}}$), and the initial gas reservoir (log$_{10}$($f_{\mathrm{corona}}$)). As discussed in \citet[][]{Blancato2019}, the three non-fitted parameters (namely, log$_{10}$($\tau$I$_{\mathrm{a}}$), $\chi_{\mathrm{out}}$, and log$_{10}$($f_{\mathrm{corona}}$)) have been shown to be relatively uninformative. Additionally, as described in \citet[][]{Rybizki2017}, observational constraints on these three parameters are less certain. Therefore, for this work we do not attempt to constrain these GCE model parameters during the MCMC routine (namely, log$_{10}$($\tau$I$_{\mathrm{a}}$), $\chi_{\mathrm{out}}$, and log$_{10}$($f_{\mathrm{corona}}$)), and instead adopt the default prior values and associated errors from \citet[][]{Rybizki2017} as our posterior parameter values and uncertainties.

Utilising the mean ([Fe/H], [Mg/Fe], [Ba/Fe], age) values for stellar populations binned in ([Fe/H], [Mg/Fe], age) space as a representative star, we model the mean abundance data across nine [Fe/H]-[Mg/Fe] cells (see Fig~\ref{mgfe}) and three different (old/intermediate/young) ages, for which the majority of our data is contained. We choose to model three independent ages within each cell because we suggest this is a better representation of a single zone GCE model, as stars formed at a particular interval of time will have undergone less gas and stellar population mixing over time. The nine cells that we model span a range in the $\alpha$-Fe plane of --0.7 $<$ [Fe/H]$<$ 0.2 to --0.2 $<$ [Mg/Fe] $<$ 0.4. The uncertainties in the [Mg/Fe], [Fe/H], and [Ba/Fe] abundances, as well as the age values, are inputted by utilising the mean error for stars in that cell, computed as the mean on the uncertainty of the measurement as reported by the GALAH catalogue. For the mean values, we utilise the method described in Section~\ref{method} to determine the mean [Mg/Fe], [Fe/H], [Ba/Fe], and age values, taking into account the uncertainties in the measurements by sampling the errors 200 times under a Gaussian distribution.
 
\subsection{Model inferences}

\begin{figure}
    \centering
    \includegraphics[width=\columnwidth]{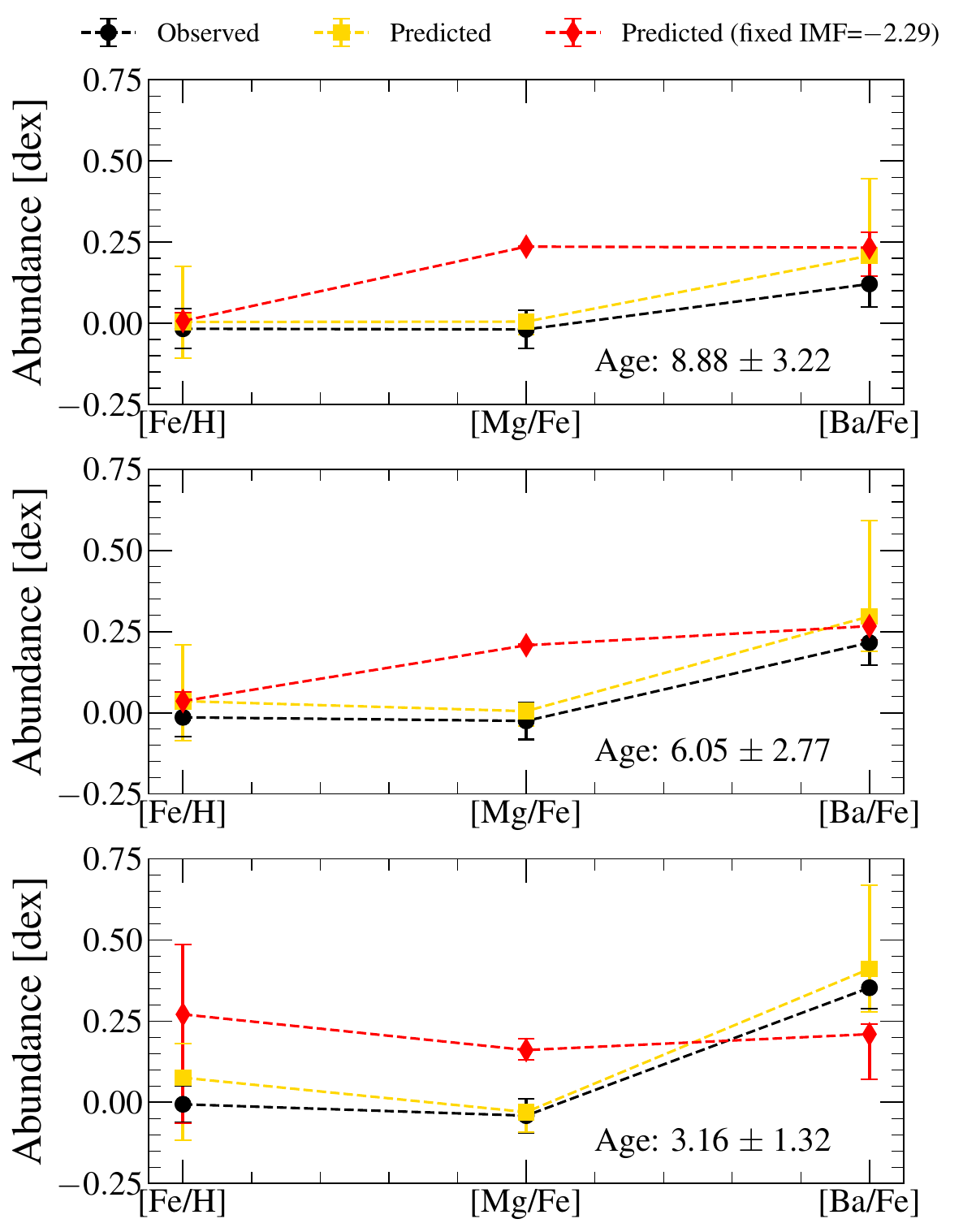}
    \caption{Observed (predicted) chemical abundances in black (yellow) for the chemical cell with --0.1 < [Fe/H] < 0.2 and --0.2 < [Mg/Fe] < 0 (i.e., cell (4,4,)) in the old (top), intermediate (middle), and young (bottom) age bins used (where the uncertainty is the mean error in age for such populations). The predicted abundances outputted from modelling the data in Chempy agree well with the chemical abundance measurements from GALAH DR3 when considering all four parameters we vary as independent. Conversely, when fitting the same abundances in the same cell under a fixed IMF (with $\alpha_{\mathrm{IMF}}$=$-2.29$, shown in red), Chempy is unable to predict abundances that match the observations.}
    \label{abundance_chempy}
\end{figure}

For each run of the fitting procedure, Chempy delivers the best fitting GCE model parameters as well as the goodness of fit of the Chempy model to the data given these optimised GCE model parameters. Fig~\ref{abundance_chempy} shows some example predictions of the elemental abundances outputted from Chempy given the best fitting posterior GCE model parameters. In Fig~\ref{abundance_chempy}, the observed mean and mean error of the chemical abundances from GALAH DR3 are illustrated in black and the predicted abundances outputted from Chempy are shown in yellow, for stars in the chemical cell defined by --0.1 < [Fe/H] < 0.2 and --0.2 < [Mg/Fe] < 0 (the same cell for which we show the posterior sample distributions in Appendix~\ref{app_mcmc}). The top panel of Fig~\ref{abundance_chempy} shows those observed/predicted abundances for stars in the old age bin (i.e., 8 < Age < 10 Gyr), the middle panel shows the results for the intermediate age bin (i.e., 5 < Age < 7 Gyr), and the bottom panel shows those in the young age bin (i.e., 2 < Age < 4 Gyr). 

It is evident that for this chemical cell Chempy is able to accurately determine and optimize the best fitting GCE model parameters to best reproduce the set of observed chemical abundances, suggestive that the model is able to fit the data. We find that Chempy's generated abundance model and GCE model parameter optimization is a good fit to the data for the majority of the chemical cells we model. We typically find agreement between the observed/predicted abundances within the 1-$\sigma$ uncertainties. However, we do note that for some cells and age bins, the models are unable to match the observed abundance of [Ba/Fe] by $\sim$0.1 dex. This is suggestive that more intricate nucleosynthetic prescriptions are required by the Chempy model in order to explain the observed [Ba/Fe] chemical abundance. In order to quantify how well different [Fe/H] and [Mg/Fe] cells are able to model the data, we compute the root-mean-squared (RMS) value of the differences between the observed and predicted abundances for every cell in Fig~\ref{rms_cells} in Appendix~\ref{app_rms_cells}.

To validate that we do not obtain similarly good fits when we fix the Chempy parameters, we show in Fig~\ref{abundance_chempy} the predicted abundances outputted from fitting the observed ([Fe/H], [Mg/Fe], [Ba/Fe], age) data within the same chemical cells, adopting a fixed value of the high mass slope of the IMF ($\alpha_{\mathrm{IMF}}$ = --2.29). It is evident that the Chempy model is unable to fit and optimize the GCE model parameters as well when fixing the $\alpha_{\mathrm{IMF}}$ parameter, indicating that the Chempy model requires this parameter to vary in order to match the observed chemical abundances.

\setlength{\tabcolsep}{14pt}
\begin{table*}
\centering
\caption{Chempy parameter priors and limits used in this work. Each MCMC run was performed using 20 walkers over 1,000 steps. The definitions of the parameters are taken from \citet[][]{Rybizki2017}.}
\begin{tabular}{ |p{2cm}|p{7.cm}|p{1.4cm}|p{1cm}|p{1.cm}}
\hline
Parameter Name & Definition &  $\theta_{\mathrm{prior}}$ & Limits & Optimized\\
\hline
\hline
$\alpha_{\mathrm{IMF}}$ & high-mass slope of the \citet[][]{Chabrier2001} IMF &--2.29\pm0.2  & [--10,10] & Yes\\
log$_{10}$(SNIa) & number of SN Ia exploding per M$_{\odot}$ over 15 Gyr & --2.75\pm0.3 & [--10,10]& Yes\\
log$_{10}$($\tau$I$_{\mathrm{a}}$) & SN Ia delay time in Gyr for \citet[][]{Maoz2010} distribution& --0.8\pm0.3 & [--5,--1]& No\\
log$_{10}$(SFE) & star formation efficiency governing the infall and ISM gas mass & --0.3\pm0.3 & [--10,10] & Yes\\
log$_{10}$(SFR$_{\mathrm{peak}}$) & peak of star formation rate in Gyr & 0.55\pm0.3 & [--10,10]& Yes\\
$\chi_{\mathrm{out}}$ &fraction of stellar feedback outflowing to the corona  &  0.5\pm0.2 & [0,1]& No\\
log$_{10}$($f_{\mathrm{corona}}$) & corona mass factor times total SFR gives initial corona mass & 0.3\pm0.3 & [--$\infty$,$\infty$]& No\\
\hline
\hline
\end{tabular}
\label{tab_chempy}
\end{table*}

\subsection{Chempy's GCE model parameters}

Proceeding from our comparison between the observed chemical abundances from GALAH with those predicted from the Chempy models, we now turn our attention towards analysing the resulting posterior distribution values of the GCE model parameters we fit. We recall that we have modelled an old, intermediate, and young sample of stellar populations in nine different chemical cells, yielding a total of twenty-seven posterior distribution values of the model parameters fitted (namely, the $\alpha_{\mathrm{IMF}}$, log$_{10}$(SNIa), log$_{10}$(SFE), and log$_{10}$(SFR$_{\mathrm{peak}}$)).

\begin{figure*}
    \centering
    \includegraphics[scale=0.25]{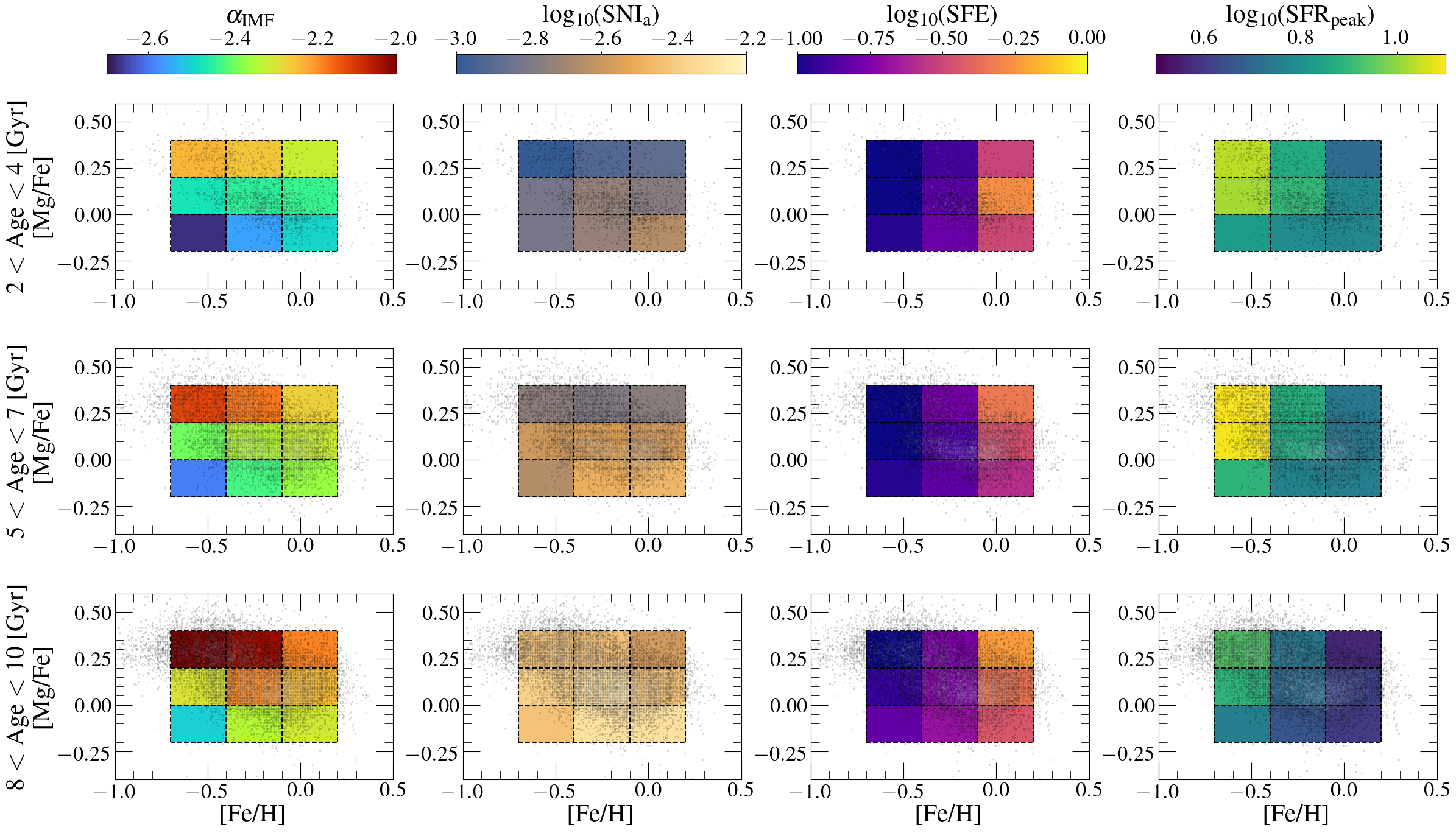}
    \caption{[Fe/H]-[Mg/Fe] distribution of the young (top), intermediate (middle), and old (bottom) disc sample. Overplotted as dashed black lines are the 9 cells we model from the 5$\times$5 chemical cell grid, each of which display the resulting $\alpha_{\mathrm{IMF}}$, log$_{10}$(SNIa), log$_{10}$(SFE), and log$_{10}$(SFR$_{\mathrm{peak}}$) values outputted from the Chempy model fits. As can be seen, the model parameters vary across [Fe/H], [Mg/Fe], and age.}
    \label{fig:new}
\end{figure*}

We summarise the Chempy parameter outputs in the chemical plane in Figure~\ref{fig:new}. This is a striking summary of the variation of each parameter across the chemical plane. There are gradients in each parameter at fixed age --across [Fe/H] and [Mg/Fe]--, and across age, at fixed ([Fe/H],[Mg/Fe]).

\subsection{Chempy's galactic parameters and correlations with stellar abundances and ages}

\begin{figure*}
    \centering
    \includegraphics[width=\textwidth]{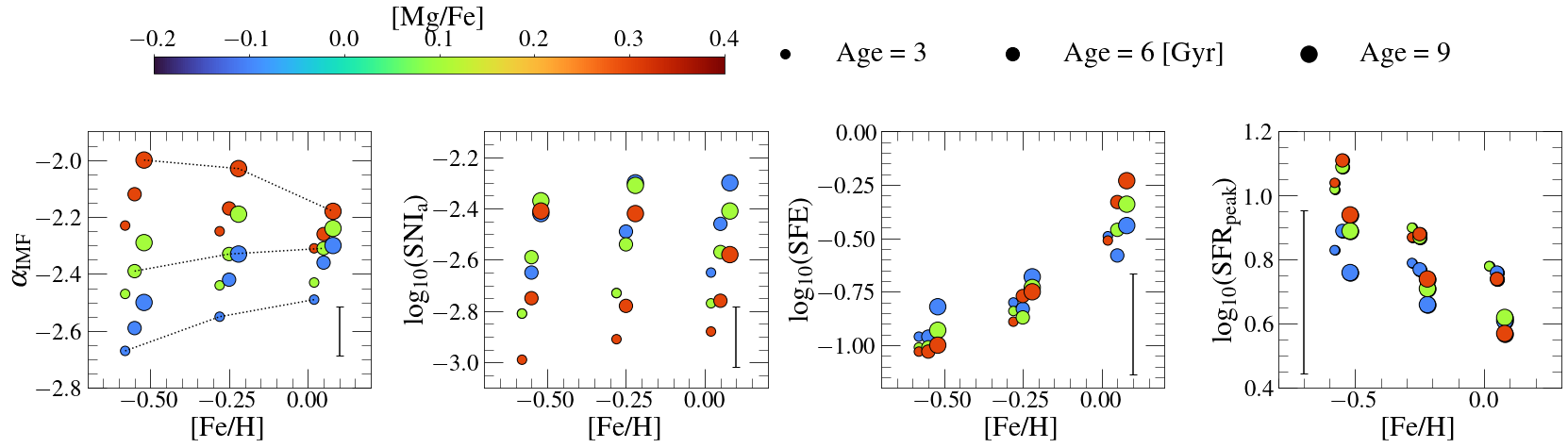}
    \caption{
    Mean posterior parameter values determined from running the flexible GCE model from Chempy on the stellar populations selected from the nine chemical cells studied as a function of [Fe/H], colour coded by [Mg/Fe], and classified by age as demarked by the legend. We note here that the [Fe/H] value is an approximate value of the mean [Fe/H] for the stellar population plotted (for illustrative purposes). It is evident that the model parameters vary across [Fe/H], [Mg/Fe], and age. The mean error bars inferred on each model parameter are indicated in the bottom corner of each panel. See text for further details.}
    \label{chempy_feh}
\end{figure*}

\begin{figure*}
    \centering
    \includegraphics[width=\textwidth]{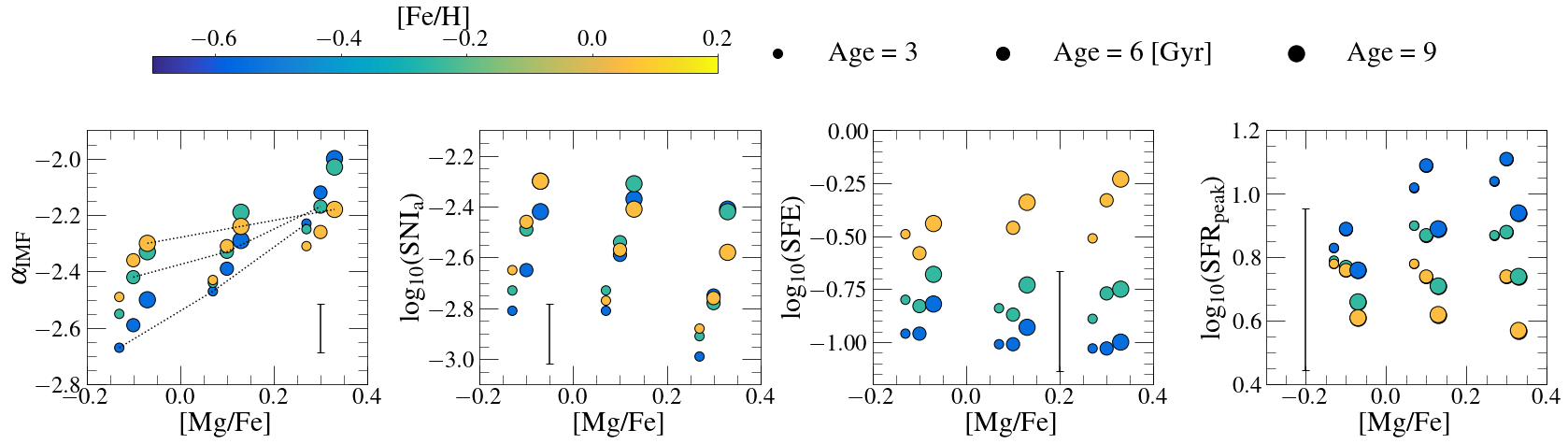}
    \caption{The same as Fig~\ref{chempy_feh}, but now as a function of [Mg/Fe] and colour coded by [Fe/H]. The [Mg/Fe] value is an approximate value of the mean [Mg/Fe] for the stellar population plotted (for illustrative purposes)}
    \label{chempy_mgfe}
\end{figure*}

To interpret the variations in the GCE model parameters seen in the maps of Fig \ref{fig:new}, we explore the correlations between these parameters and [Fe/H], [Mg/Fe], and age. Fig~\ref{chempy_feh} shows how the resulting posterior values for the GCE model parameters fitted in Chempy for different chemical cells vary systematically with [Fe/H], [Mg/Fe], and age. The size of the circle marker in Fig~\ref{chempy_feh} indicates the age of the sample being illustrated, as noted in the legend. Here, each point is coloured by the median [Mg/Fe] abundance ratio of their respective chemical cell, as marked by the colour bar. We note that the [Fe/H] values displayed are not the true mean values of the chemical cells, but are the median values within the [Fe/H] range of those cells, to allow for easier data visualisation. Fig~\ref{chempy_mgfe} shows the same relation but now as a function of [Mg/Fe] and colour coded by [Fe/H].

Together, Figures \ref{chempy_feh} and \ref{chempy_mgfe} are demonstrative of the ordered change in some of the underlying GCE model parameters that are consistent with the data in a one-zone model framework. Focusing in more detail on each model parameter independently, we report the following: 

 $\bullet$  For the $\alpha_{\mathrm{IMF}}$ parameter, we find an interesting difference when examining how the $\alpha_{\mathrm{IMF}}$ slope value varies around [Mg/Fe]. For different [Mg/Fe], we see that the high-mass end of the IMF slope changes significantly at fixed age and [Fe/H]. We note that this difference is strikingly large, ranging from a value of $\alpha_{\mathrm{IMF}}$ $\sim$--2.7 at the most [Mg/Fe]-poor cell to a value of $\alpha_{\mathrm{IMF}}$ $\sim$--2.2 at the most [Mg/Fe]-rich one (for an age of $\sim$3 Gyr and [Fe/H]$\sim$--0.6). We also find that the variation of $\alpha_{\mathrm{IMF}}$ as a function of [Mg/Fe] is steeper for the most [Fe/H]-poor cells, where the difference at fixed age can range up to $\sim$0.6. Conversely, for the most [Fe/H]-rich cells, this difference is much smaller, on the order of $\sim$0.3. Furthermore, when pivoting around age at fixed [Fe/H] and [Mg/Fe], we find that the older populations display overall a higher $\alpha_{\mathrm{IMF}}$ value than their younger counterparts. Here, the older populations have, on average, a $\sim$0.1 higher value than younger populations of the same [Fe/H] and [Mg/Fe]. In a similar vein, we also see a positive correlation of $\alpha_{\mathrm{IMF}}$ slope with [Fe/H] across chemical cells with [Mg/Fe] < 0.2, where those cells that contain the most metal-poor populations have a more bottom-heavy IMF for fixed age and [Mg/Fe] than more metal-rich counterparts. Conversely, we find that for the most [Mg/Fe]-rich cells (namely, [Mg/Fe] > 0.2), this positive slope turns into a negative one with decreasing [Fe/H] and fixed age. At this [Mg/Fe] value, the most [Fe/H]-poor cell displays $\alpha_{\mathrm{IMF}}$ $\sim$--2 and the most [Fe/H]-rich a value of $\alpha_{\mathrm{IMF}}$ $\sim$--2.2. However, within the uncertainties, the differences in the $\alpha_{\mathrm{IMF}}$ with [Fe/H] are negligible for the intermediate [Mg/Fe] cells. In a similar vein, we also note that the mean of the values obtained for the exponent of the high-mass slope of the IMF, across the cells, is consistent with \citet[][]{Chabrier2003} ($\alpha_{\mathrm{IMF,Chabrier}}$ = --2.3\pm0.3) and \citet[][]{Salpeter1955} ($\alpha_{\mathrm{IMF,Salpeter}}$ = --2.3). In summary, we find that at fixed age and [Fe/H], chemical cells with higher [Mg/Fe] have a more top-heavy IMF than their lower-[Mg/Fe] siblings (in terms of the high-mass end of the IMF slope). We find that within the uncertainties, the $\alpha_{\mathrm{IMF}}$ parameter does not vary with [Fe/H] (for fixed [Mg/Fe] and age) and age (for fixed [Fe/H] and [Mg/Fe]).

$\bullet$ For the log$_{10}$(SNIa) parameter, we find that, within the uncertainties, it has no dependence on [Fe/H] or [Mg/Fe]. However, we do find that it depends on age. At fixed [Mg/Fe] and [Fe/H], we find that older stellar populations have a higher contribution of SNIa than their younger counterparts. This is likely due to the delayed onset of SNIa.

$\bullet$ When examining the values of the log$_{10}$(SFE) parameter, we see that this parameter depends strongly on [Fe/H] and negligibly on [Mg/Fe] and/or age. We find that stellar populations of the same age and [Mg/Fe] in more metal-poor cells display lower SFE values than their metal-rich counterparts, although we note that within the uncertainties, this relation becomes much weaker. Furthermore, for a given chemical cell, intermediate populations require the lowest SFE value and older populations require the highest. 

$\bullet$ For the log$_{10}$(SFR$_{\mathrm{peak}}$) parameter, we report that more [Fe/H]-poor populations are fit on average by a model with a higher peak star formation rate than the [Fe/H]-rich counterparts. At fixed [Fe/H] and age, higher [Mg/Fe] populations have higher log$_{10}$(SFR$_{\mathrm{peak}}$). Within a chemical cell, younger populations have higher log$_{10}$(SFR$_{\mathrm{peak}}$) parameter values. However, we find that the uncertainty on the values of this model parameter are too large to make any conclusive statements.

\subsection{Chempy's GCE model parameters and correlations with stellar orbits}
\label{chempy_orbits}
We have seen clear correlations between the GCE model parameter values and [Fe/H], [Mg/Fe], and age. This result calls for the exploration of these relations in more detail in terms of the orbital properties of the stars. We use the orbital radius\footnote{Computed as ($R_{\mathrm{apo}}$ + $R_{\mathrm{peri}}$)/2,  where $R_{\mathrm{apo}}$ and $R_{\mathrm{peri}}$ correspond to the apogalacticon and perigalacticon values of each stars orbit, respectively.} and mean vertical height above the Galactic disc (i.e., $z_{\mathrm{max}}$) for old/intermediate/young stellar populations in each chemical cell, in order to examine if the posterior values of the GCE model parameters fitted vary with average position in the Galaxy. We note that the orbital radius quantities were calculated utilising the same sampling chemical cells method used to determine the [Ba/Fe]-age relation values (i.e., by sampling 200 times the [Fe/H]-[Mg/Fe] plane and determining the mean for every cell and age bin).

\begin{figure*}
    \centering
    \includegraphics[width=\textwidth]{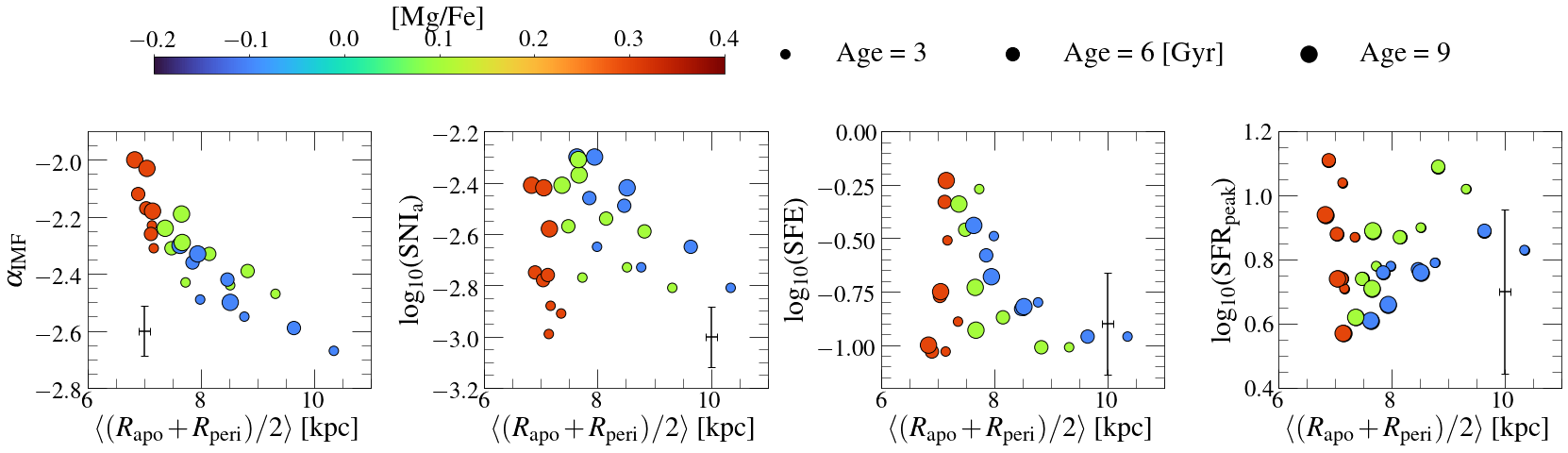}
    \caption{The same values from Fig~\ref{chempy_feh}, but now as a function of mean orbital radius and colour coded by mean [Mg/Fe]. There is a clear link between the mean posterior values of the model parameters and the mean orbital radius of a stellar population. See main body of text for more details.}
    \label{chempy_rmean_fixedfeh}
\end{figure*}

Fig~\ref{chempy_rmean_fixedfeh} shows the same results from Fig~\ref{chempy_feh} but with respect to the 
orbital radius for old/intermediate/young stars in every chemical cell. Similarly to when examining the resulting posterior GCE parameter values with respect to ([Fe/H],[Mg/Fe],age), we find correlations between each model parameter and $\langle$($R_{\mathrm{apo}}$ + $R_{\mathrm{peri}}$)/2$\rangle$, where it is the $\alpha_{\mathrm{IMF}}$ parameter that shows the strongest correlation. These are far less structured in orbital space compared to the trends with the [Fe/H], [Mg/Fe] and age, with the notable exception of the $\alpha_{\mathrm{IMF}}$ parameter.

Figure \ref{chempy_rmean_fixedfeh}  shows that for stars of a given age and [Mg/Fe], those populations with larger orbital radius values (i.e., further away from the Galactic centre and up to $\sim$10kpc) show a more bottom-heavy $\alpha_{\mathrm{IMF}}$ value ($\sim$--2.7) than those stellar populations closer to the inner Galaxy ($\sim$7kpc and $\alpha_{\mathrm{IMF}}$ $\sim$--2.1). This result is also observed as a trend with [Mg/Fe], where more [Mg/Fe]-rich populations present a more top-heavy $\alpha_{\mathrm{IMF}}$ value than their lower [Mg/Fe] counterparts. Furthermore, the scatter around the mean trend is consistent with the size of the error bar on the IMF parameter, indicative of the tightness of this relation. This finding is suggestive of a possible spatially (radially) varying high-mass slope of the IMF in the Milky Way. Along those lines, for fixed age and [Fe/H], those stellar populations that orbit closer to the Galactic centre on average endure a higher number of SNIa contributions than stellar populations at larger mean Galactocentric distances. With regards to the SFE value, we find no clear relation between $({R_{\mathrm{apo}}+R_{\mathrm{peri}}})/2$, [Mg/Fe] and age. We find that stellar populations at fixed orbital radii show a range in SFE across all ages, for which the scatter becomes smaller at larger orbital radii. In terms of the peak of the star formation rate, we find that on average, stellar populations that are older have lower log$_{10}$(SFR$_{\mathrm{peak}}$) values than their younger, irrespective of [Mg/Fe]. However, we note that for this model parameter, the uncertainties are too large to make any quantitative conclusion.

In summary, in this Section we have set out to model our observational measurements of [Fe/H], [Mg/Fe], [Ba/Fe] for old, intermediate, and young stellar populations in nine different chemical cells ranging in [Fe/H] and [Mg/Fe]. In doing so, we have been able to estimate the best fitting posterior distribution values for key parameters governing the Chempy GCE model, those being the high-mass slope of the IMF ($\alpha_{\mathrm{IMF}}$, the number of SNIa explosions (log$_{10}$(SNIa)), the SFE (log$_{10}$(SFE)), and the peak in the star formation rate (log$_{10}$(SFR$_{\mathrm{peak}}$)), as well as study their relation with respect to [Fe/H], [Mg/Fe], age, and (average) position in the Galaxy. 
The results presented in this Section, which show a smooth variation of some of the Chempy GCE model parameters with age, chemical compositions, and average orbital radii, suggest that the chemical evolution of the Galaxy can be described by GCE model parameters that change systematically with birth place and time.  In the next section, we will frame our findings in the context of current work, and discuss the implications of our results with respect to our current understanding of the formation of the Milky Way.

\section{Discussion}
\label{discussion}

\subsection{Age-abundance profiles of neutron capture elements in the Galactic disc}

Across the well studied [Fe/H]-[$\alpha$/Fe] plane, chemical cells (or ``mono-abundance populations (MAPS)" (\citealp[e.g.,][]{Bovy2012,bovy2012b, Martig2016, Mackereth2017, Mackereth2019_disc})) have set a new course for exploration in many avenues of Galactic archaeology. The ability to select stellar populations with the same number of SNII and SNIa contributions is also, however, of particular utility from a nucleosynthetic standpoint to study chemical elements synthesized in independent processes. In this paper we use chemical cells to study the age-abundance relations of the $s$- and $r$-process neutron capture elements, of [Ba/Fe] and [Eu/Fe], respectively, across a grid in [Fe/H]-[Mg/Fe] space. Within cells, this illuminates the relationship between SNII/SNIa and neutron capture yields over time. Between cells, the age-abundance profiles are an imprint of the architectural assembly of the galaxy, as ([Fe/H], [Mg/Fe], age) are a direct link to birth radius (\citealp[e.g.,][]{Frankel2018,Minchev2018,Ness2019}), thus linking the neutron capture and SNII/SNIa  nucleosynthetic sources over space and time in the Galaxy. 

  Section~\ref{results} maps the age-abundance relations for [Ba/Fe] and [Eu/Fe], across the [Fe/H]-[$\alpha$/Fe] plane. The overall mean relations between age-[Ba/Fe] and age-[Eu/Fe] reported in this study broadly match that from previous work focusing both on stellar populations in the field  (\citealp[e.g.,][]{Dasilva2012,Reddy2015,Nissen2016,Nissen2017,Bedell2018, Hayden2020, Griffith2021_galah, Zinn2021}) and in open/globular clusters (\citealp[e.g.,][]{Dorazi2009,Maiorca2011,Magrini2018}). These suggest that younger stellar populations manifest on average higher contributions from the $s$-process nucleosynthetic channel (although possibly not in the same amounts for different $s$-process elements, see \citealt[][]{Dorazi2017} and \citealt[][]{Reddy2017}).
  
In stepping through individual chemical cells, we see remarkably structured age-[Ba/Fe] relations, that vary neatly across in slope and amplitude. The age-[Ba/Fe] relations show a negative slope across all cells for stars younger than 10 Gyr, of approximately $\sim$ $-$0.05 dex/Gyr (i.e. in agreement with previous work \citealp[e.g.,][]{Bedell2018}). Conversely, for the oldest ages ($>$ 10 Gyr) we see an increase in the [Ba/Fe] value.  This could be due to either or the combination of two factors: i) the difference in the onset of AGB-wind and SNIa contributions, leading to a high primordial Ba that decreases with Fe production until onset of AGB-winds; ii) an issue in the GALAH [Ba/Fe] determination for the oldest RGB stars (see \citet[][]{Hayden2020} for a discussion).
We see that the abundance ratio between the $r$-process element Eu, and the iron-peak element Fe, is constant with age within each of our chemical cells. However, the mean [Eu/Fe] value changes primarily as a function of [Mg/Fe], but also to some extent in [Fe/H]. 

Furthermore, we note that the high enhancement in [Ba/Fe] at young ages has been argued to be an artifact of the LTE approximation in making the [Ba/Fe] measurement (see \citet[][]{Reddy2017}). However, other $s$-process elements (e.g. Ce, La) show qualitatively similar gradients, albeit reaching lower overall enhancements at young ages, with correspondingly flatter gradients (\citealp[][]{Bedell2018,Baratella2021}). This is indicative of a chemical enrichment mechanism that decreases the $s$-process abundances with respect to iron over time (also see \citet[][]{Banerjee2018} and \citealt{Prantzos2018}).

The correlations between spatial location and the resulting [Fe/H]-[$\alpha$/Fe] abundances of stars in the Galactic disc have emerged in the past few decades thanks to the large number of spectroscopic studies (\citealp[e.g.,][]{Bensby2014,Hayden2015,Bovy2016,Mackereth2017}). Stars with higher [$\alpha$/Fe] are concentrated to the inner regions of the Galaxy, at larger heights from the plane, and stars with lower [$\alpha$/Fe] to the outer Galaxy, and nearer to the mid-plane. Furthermore, there is an overall decreasing mean [Fe/H] with Galactic radius. Our finding on the different age-[Ba/Fe] relations observed for different stellar populations in the Galactic disc selected in chemical cells of [Fe/H] and [Mg/Fe] is presumably a chemical signature of the relationship between the orbital architecture of disc stars and their birth place in the Galaxy (given that [Fe/H], [Mg/Fe], and age link to birth radius). In Section~\ref{results_radius} we see that within some chemical cells, there is a gradient of mean orbital radius along the age-[Ba/Fe] relations. Moreover, across the chemical cells studied, there is a gradient of mean orbital radius at fixed age with [Fe/H] and [Mg/Fe]. It is non-trivial to interpret these results; they involve correlations across multiple dimensions of (Fe,Mg,Ba,age) and the orbital radius values are subject to change over time due to radial migration \citep[e.g.][]{Minchev2018, Frankel2018, Frankel2019, Frankel2020}. Qualitatively, however, we can consider that from a nucleosynthetic standpoint, the higher [Ba/Fe] abundances at lower ages for stars at fixed ([Fe/H]-[Mg/Fe]) are not directly governed by the total number of SN contributions. Presumably, the [Ba/Fe] abundances within cells are a direct consequence of where those stellar populations are positioned (and, moreover, born) in the Galactic disc. Thus, at each fixed [Fe/H]-[Mg/Fe] and age bin, we have a group of stars born at a particular time, and place - as informed by their [Ba/Fe] abundance. The present-day correlations that we observe with orbital radius, and the smooth variation of the correlations across the [Fe/H]-[Mg/Fe] plane, indicate the assembly of the disc was ordered. Furthermore, under the assumption of metallicity dependent Ba yields, at fixed [Fe/H], the nucleosyntheic consequences are striking in that our results suggest that different regions of the Galactic disc host a different total number of low-/intermediate-mass AGB stars (at fixed SNII/SNIa contributions and age). The correlations we measure indicate fewer low-/intermediate-mass AGB stars in the inner regions of the Galactic disc, and a higher ratio in the outer regions. Quantitatively, we can use GCE modelling to reduce these variables to a set of underlying model parameters that describe the chemical evolution of the Galaxy. 

We note that recent work \citep[e.g.][]{Zhang2021} suggests that high [Ba/Fe] ratios for young $\alpha$-rich stars is consistent with the stars being old, and products of binary evolution. However, if we exclude the high-$\alpha$ stars from our analysis ([Mg/Fe] $>$ 0.2, rows 3 and 4 in Figure 5), our results and conclusions do not change.

\subsection{Modelling stellar populations in the Galactic disc with Chempy}

In contrast to some of the most recent work (\citealp[e.g.,][]{Spitoni2021,Johnson2021}), in this paper we attempt to model chemical abundances over time in the Milky Way disc using the simple and flexible Chempy GCE model (see also \citealt[][]{Philcox2019}). Briefly, the method employed in this paper deviates from more classical approaches in three main ways: i) we employ narrow ranges of [Mg/Fe], [Fe/H], and age, in order to study [Ba/Fe] at different times, for populations with very similar SNII/SNIa contributions. In doing so, we conduct an analysis that enables us to connect the observational abundance data (Fe,Mg,Ba) and age to chemical evolution theory. Essentially we can then ask, how and why does one element abundance vary over time in a small neighbourhood of chemical sub-space; ii) because we model stellar populations in chemical cells in bins of stellar age, our approach is suitable for a one-zone model framework. An alternative approach, for example, is to combine stars of different ages across large regions in chemical space (for example, the high- and low-$\alpha$ sequences),  and assume different zones of evolution. Indeed our biggest restriction rests upon our assumption of a one-zone model framework; the parameterisation that we find is of course tied to this model description. However, because we are modelling stellar populations in bins of stellar age, we believe that this assumption is well motivated; iii) we ultimately have very few assumptions; the underlying set of parameters that describe the GCE model (namely, $\alpha_{\mathrm{IMF}}$, log$_{10}$(SNIa), log$_{10}$(SFE), and log$_{10}$(SFR$_{\mathrm{peak}}$)) are allowed to vary to find the best fit to the data.

Under our assumptions, we find that our model can reproduce the abundance measurements (Fig~\ref{abundance_chempy}). The results of our modelling of the observed [Fe/H], [Mg/Fe], [Ba/Fe], and age of twenty-seven different stars (representing an underlying stellar population) in the Galactic disc (at three ages) are as follows: 

 i) There is a smooth and gradual change in the estimated values of all the GCE model parameters fit for the different stellar populations across chemical cells that varies with age, [Fe/H], and [Mg/Fe]. However, the uncertainties on the log$_{10}$(SFE) and log$_{10}$(SFR$_{\mathrm{peak}}$) parameters are large and restrict us from being able to make any quantitative constrain on any possible relation. Given our results, it is likely that the Chempy model is primarily only able to constrain the $\alpha_{\mathrm{IMF}}$ and log$_{10}$(SNIa) parameters.
 
ii) Within the uncertainties, the high-mass slope of the IMF parameter varies with [Mg/Fe], but not with [Fe/H] or age. At lower [Mg/Fe] the IMF is more bottom heavy, and vice-versa at higher [Mg/Fe].

iii) At fixed [Fe/H] and [Mg/Fe], older populations have systematically smaller $\alpha_{\mathrm{IMF}}$ parameter values than younger populations. However, this difference is negligible within the uncertainties. 

iv) The SNIa parameter shows a weak [Mg/Fe] dependence, with a decrease to more negative values at higher [Mg/Fe].

v) The SFE increases significantly as a function of [Fe/H], with the scatter at a fixed [Fe/H] being smaller than the uncertainty on the measurement (across all [Mg/Fe] and age).

vi) The peak of the SFR decreases as a function of [Fe/H], with the scatter at a fixed [Fe/H] being smaller than the uncertainty on the measurement (across all [Mg/Fe] and age).
 
vii) The IMF high-mass slope parameter shows a tight correlation with mean orbital radius. The scatter around the mean of this orbiting radius-$\alpha_{\mathrm{IMF}}$ relation is of the order of the uncertainty of this parameter (as seen from Figure~\ref{chempy_rmean_fixedfeh}). Stars around the solar neighborhood have a more top-heavy $\alpha_{\mathrm{IMF}}$ value.

Overall, stars in the inner Milky Way  have a more top-heavy $\alpha_{\mathrm{IMF}}$, a higher number of SNIa contributions, and a range of star formation efficiencies and peak star formation values (the latter two are dependent on age and [Fe/H]). 
 
An important question that falls out explicitly from our procedure is: by fitting the data with our GCE model, do we learn about the model or about the data, or both? With regards to the model, we have learned that we can satisfactorily fit the set of observed chemical abundances under a variable set of GCE model parameters. This suggests that the yield tables used likely work reasonably well except for the lowest metallicity regime. This hypothesis is supported by the higher root-mean-squared value obtained for more [Fe/H]-poor cells when comparing the observed and predicted ([Fe/H],[Mg/Fe],[Ba/Fe]) abundances (see Fig~\ref{rms_cells} in Appendix~\ref{app_rms_cells}). With regards to the data, we have learned that the GCE model parameters vary with the physical properties of the stars (namely, [Fe/H], [Mg/Fe], [Ba/Fe], age, $(R_{\mathrm{apo}}+R_{\mathrm{peri}})/2$, $z_{\mathrm{max}}$). This suggests that there is an intimate link between the physical properties of stars from the data with the underlying chemical evolution model. Given all our findings, we discuss the implications of these results in the context of other Galactic chemical evolution models approaches and results further in the following subsection.

\subsection{Our results in the context of Galactic chemical evolution models}

The question of how the Milky Way has chemically evolved throughout cosmic time and reached its current form has been a longstanding and non-trivial question to answer. In order to tackle this, many previous works have sought to interpret the chemical abundance distribution of stellar populations in the Galaxy via the use of Galactic chemical evolution models. Beginning with the pioneering work by \citet[][]{Tinsley1980}\footnote{and her amazing work between 1940-1980.}, there have been many papers that have attempted to decipher how the Milky Way disc has reached its current form and chemical compositions using a plethora of different chemical evolution models (\citealp[e.g.,][]{Matteucci1989,Rana1991,Chiappini1997,Travaglio1999,Kobayashi2006,Schonrich2009, Minchev2013,Pilkington2012,Nomoto2013, Kobayashi2020b,matteucci2021modelling}). 

In order to explain abundance-age relations for Ba, GCE models have been proposed suggesting Ba is primarily released in the winds of low/intermediate mass AGB stars (\citealp[e.g.,][]{Maiorca2012,Cristallo2015,Karakas2016,Choplin2018,Rizzuti2019,Kobayashi2020b}). Under this assumption, it is plausible to conceive that those lower-mass and more long-lived AGB stars have polluted the interstellar medium (ISM) with higher doses of Ba that younger generations of stars used in order to form, resulting in the increased $s$-process abundances observed for these populations. However, of particular importance is the novel result obtained in this work: the negative age-[Ba/Fe] relations for Galactic disc populations persist (and vary) for stellar populations selected in a SNII and SNIa reference frame. Under the assumption that Ba is primarily released in the winds of low/intermediate AGB stars, our findings from Fig~\ref{bafe_cells} at face value suggest that there is a different total number of low-/intermediate-mass AGB stars in Milky Way disc populations selected based on their ([Fe/H], [Mg/Fe], age) values. 

The number of low-mass stars present in a stellar population is directly linked to the IMF. Although there is a general consensus for a universal IMF amongst the Galactic community (\citealp[see ][and references therein]{Bastian2010}), in recent years there have been several works that argue for variations in this key GCE parameter for different Galactic and extra-Galactic stellar populations (\citealp[][]{Ballero2007, Brandner2008, Espinoza2009, Lu2013, Hallakoun2020}), suggesting that the universally claimed value of the IMF should be revised to accommodate observational results of low-mass stars \citep[][]{Romano2005}, or that it should be considered to vary in a more complex manner \citep[][]{Guszejnov2019}. Variations in the IMF have also been postulated given integrated field spectroscopic results for stellar populations in external galaxies (\citealp[e.g.,][]{Hoversten2008,Treu2010,Conroy2012,Cappellari2012,Cappellari2013,Ferreras2013,Lyubenova2016,Davis2017,Vandokkum2017, Parikh2018,Dominguez2019}). Our findings on the link between orbital radius and the age-[Ba/Fe] relations of different Galactic disc populations selected in chemical cells of [Mg/Fe] and [Fe/H] therefore do not seem totally unreasonable given these recent contentious results. However, before arriving to any conclusion based on our findings, it is pivotal to discuss our empirical results in the context of Galactic chemical evolution models.

The IMF has been explored previously within the Milky Way: \citet[][]{Chiappini2000} attempted to impel varying IMF models to try and reproduce the observed chemical distribution of stars in the field of the Galaxy, reaching the conclusion that, under the assumption of the two-infall model \citep[][]{Chiappini1997}, a universal IMF yielded a better prescription to the data. Conversely, in a more recent study \citet[][]{Guszejnov2019} utilised a range of simulated galaxies (including MW-like galaxies) and implement GCE models with varying IMF. The results from \citet[][]{Guszejnov2019} showed that no IMF models currently in the literature were able to reproduce the claimed IMF variation in early-type galaxies (ETGs) and/or dwarf galaxies without severely violating observational constraints in the MW, suggesting that a more complex form of the IMF is required. These two examples clearly illustrate how much we still have to learn about this fundamental GCE model parameter.

Recent work has set out to interpret the chemical evolution of the Galaxy by either studying the ratio of elemental abundances in stars whose element genesis are well constrained (\citealp[e.g.,][]{Weinberg2019, Griffith2021,Griffith2021_galah}), or by invoking multi-zone GCE models (\citealp[e.g.][]{Johnson2021}). In more detail, \citet[][]{Weinberg2019} showed that by studying the ratio of elemental abundances in different Galactic components with respect to [Mg/H] instead of the usual [Fe/H], a deeper insight into the chemical evolution of the Galactic disc could be reached, and suggested that the chemical compositions of the high- and low-$\alpha$ sequences observed in the MW are independent of spatial location. Following this study, \citet[][]{Griffith2021} showed that this result could be expanded, to a lesser extent, when comparing the chemical compositions of high- and low-alpha sequences in the Galactic bulge. However, the authors also suggested that the differences in the chemical compositions of bulge and disc populations could be linked to a small, yet noticeable, variation in the exponent of a power-law IMF of the Galaxy (on the order of $\lesssim$0.3). Along those lines, the results presented in \citet[][]{Johnson2021} revealed that, under the framework of a hybrid multi-zone GCE model that accounts for stellar migration (informed by cosmological simulations) and a universal IMF, the characteristic $\alpha$-bimodality observed in the MW disc could not be reproduced; the authors suggested that more dramatic evolutionary pathways are required. Conversely, results from cosmological simulations seem to arrive to the $\alpha$-bimodality under a universal Chabrier IMF \citep[][]{Grand2018}.

Given that in our simple modelling framework we can describe the abundance data, the next natural step is to ask what other models can describe these data. We anticipate that simultaneous fitting of many elements over time using flexible multi-zone GCE models, that allow for particular parameters in the model to be fixed/free \citep[e.g.,][]{Philcox2019}, will be a very promising avenue for future exploration.

\subsection{Caveats and assumptions in the models}

In this work, we have modelled the chemical abundance data with a flexible one-zone GCE model, which in turn has enabled us to quantitatively assess the chemical abundance data of Galactic disc populations. However, it is critical to recognise that we are restricted by the assumptions made in our model. It therefore behooves us to discuss qualitatively alternative theoretical explanations for our observational results. As stated in Section~\ref{sec_bafe}, our results indicate that it is not simply over time that low-mass stars contribute in greater amounts Ba to older stellar populations. Rather, at a fixed chemical sub-space in the SNII/SNIa reference frame (i.e., [Fe/H]-[Mg/Fe]), stars of different ages have either different enrichment pathways or underlying GCE model parameters that enable the same ([Mg/Fe],[Fe/H]), but different [Ba/Fe] mean values to be reached. Moreover, across the SNII/SNIa reference plane, the mean value and slope of the age-[Ba/Fe] relations vary smoothly, with mean [Ba/Fe] decreasing at fixed age across [Fe/H] and increasing across [Mg/Fe]. 

However, two important caveats in our model are: $i$) the lacking ingredient and framework that allows stellar and gas exchange between different model zones $ii$) a possibly missing delay time parameter that enables the onset of the initial star formation at later times for stars formed in the outer regions of the disc (presumably at a later time under inside-out formation)."

The flexible one-zone model we use enables us to examine the GCE model parameters that best explain the data. However, if the trends we measure empirically from the data turn out to be a direct consequence of, and are dominated by, the process of gas mixing/exchange across the MW disc, then we need to employ a flexible multi-zone framework moving forward (\citealp[e.g.,][]{Philcox2019,Johnson2021}). Note, we specifically refer to the utility of a multi-zone model of chemical enrichment that can move gas between any two zones within a flexible time frame --as per the VICE approach \citep[][]{Johnson2021}-- which includes a prescription of radial migration. However, we propose that to test over the full parameterisation of the star formation history of the Milky Way, this modelling framework should also include the flexiblity of Chempy, so as to enable an understanding of the role of the variation of the different environmental parameters. Presumably, the high dimensional data and stellar ages will enable us to discriminate within the model in solving how data inform quantitative enrichment history. The data indicate that the underlying variance in the chemical abundance space is smoothly distributed across the [Fe/H]-[Mg/Fe] plane. This suggests that this may be tied to stellar birth location and or/time of formation. We assess, qualitatively, if our observations could be explained simply by gas exchange in an galactic disc that has a decreasing star formation rate with radius, and which forms inside-out with a decreasing metallicity gradient. We ask ourselves: "\textit{Could the smooth trends we observe in Fig~\ref{bafe_cells} arise due to gas exchange between zones (across Galactic radii) in the presence of both gas and stellar migration?}". We expect that, on average, the oldest stars are born in the inner disc, which one might assume has experienced the fastest star formation rate, whilst the youngest stars form in the outer regions \citep[i.e., inside-out formation][]{Minchev2017}. Presumably this imputes the oldest stars - on average - with lower [Ba/Fe] at fixed [Fe/H]. At the same time, the metallicity gradient is negative from the oldest to youngest stars, so gas exchange from different radii would feed (on average) younger star forming regions more metal-rich gas and older star forming regions more metal-poor gas. This could potentially enable the same ([Fe/H]-[Mg/Fe]) value to be reached in young and old populations under different star formation rates that give rise to different [Ba/Fe] ratios; whereby the older stars in the inner region might have higher [Ba/Fe], and the younger star formed in the outer region would have lower [Ba/Fe] (driven by the relationship between star formation rate and metal-rich/metal-poor gas exchange). Given that we can model the data with a one-zone model allowing parameters like the $\alpha_{\mathrm{IMF}}$, log$_{10}$(SNIa), log$_{10}$(SFE), and log$_{10}$(SFR$_{\mathrm{peak}}$) to vary, the logical next step is to move to a more flexible model to assess how and if these vary under gas exchange across radius \citep[][]{Johnson2021}.

In a similar vein, in this work we have fitted 3 data points with 4 free parameters. Given the large uncertainties obtained for the SFE and the SFR parameters, it is likely that we essentially reproduce the priors for these (i.e. the data do not constrain these, and only the $\alpha_{\mathrm{IMF}}$ and the log$_{10}$(SNIa) parameters are being shifted and determined more
precisely than from the prior). The model
marginalises over these parameters and at the same time they have the physically reasonable correlations with the other GCE model parameters. The main leverage of the model are the fits to the $\alpha_{\mathrm{IMF}}$ and log$_{10}$(SNIa) parameters. These
fix the [Mg/Fe], as the more top heavy the $\alpha_{\mathrm{IMF}}$ parameter, the more high mass SNII explosions and Mg being synthesized, and thus in order to reach the correct [Mg/Fe] ratio, the more SNIa explosions and Fe production. This is likely why both the $\alpha_{\mathrm{IMF}}$ and log$_{10}$(SNIa) are positively correlated. Along those lines, the log$_{10}$(SFE) parameter can still shift the [Fe/H] ratio slightly, but it likely does not
force the log$_{10}$(SFE) parameter to reduce its prior range by a significant amount. Thus, the
correlation of $\alpha_{\mathrm{IMF}}$ and log$_{10}$(SNIa) parameters usually settle where the [Fe/H] ratio is
in line with the data. Conversely, the log$_{10}$(SFR$_{\mathrm{peak}}$) parameter is quite complex, as it
forces the chemical enrichment to be stronger earlier on so that the [Fe/H]-dependent yields are introduced to the models in a slightly different manner. In other words: the total yields have different fractions of high- and low-[Fe/H] yields incorporated which can actually induce
specific changes in the abundance patterns, but these are much weaker
than the general SNII and SNIa trends. As a result, the log$_{10}$(SFE) and log$_{10}$(SFR$_{\mathrm{peak}}$) model parameters are
still at work, but to a lesser degree. Furthermore, although the predictions obtained yield good agreement between the observed and predicted abundances when using the default set of nucleosynthetic yields assumed by the Chempy model, it may be possible that different yield sets are a better match to the data. We note that \citet[][]{Rybizki2017} tested this hypothesis when comparing the observed abundances of a proto-star, Arcturus, and B-stars with their default yield set, and an alternative one (see their Fig 14). The authors found that, Mg and Fe specifically, were better predicted using the set of default yield tables in the Chempy model when compared to the alternatives from the literature. Thus, while it still may be possible that a unique combination of yield sets could lead to better agreement between the observed and predicted abundances, given the good agreement found with the default Chempy yield set, we are confident that the Chempy model is able to yield reliable results.

In a final remark, every analysis choice involves a trade-off between accuracy and complexity. The one-zone model approximation is somewhat in detail inaccurate, but it is simple; one could build a radial mixing model for the gas \citep[e.g.,][]{Johnson2021} and then do better (in principle), but at the cost of making more assumptions and having more parameters. We effectively use Chempy as it was intended, to model individual stars. However, a key insight we make is that we can divide the chemical abundance space up, so as to use the many stars in large surveys to obtain a set of “representative stars” of common birth environment (radius and age) and thereby explore the GCE model parameters that are consistent with a large chemical space of the data.
One-zone models are incomplete, in detail, to model together simultaneously stars born at different birth radii and with different ages, because there is probably also radial mixing of gas and supernova and AGB ejecta. However, one-zone models are straightforward to use and examine when modelling stellar populations of similar age, without the need of building a gas-exchange mixing model. Here, we model stellar populations binned in ([Fe/H], [Mg/Fe], age) space, employing representative stars, individually – one at a time – from presumed similar birth radii and age. This assumption allows us to make appropriate use of such a one-zone model framework.

\section{Conclusions}
\label{conclusion}

In this paper we have utilised the latest GALAH DR3 and \textit{Gaia} EDR3 catalogues in order to statistically study the abundance-age relations of neutron capture elements in multiple chemical cells across a [Fe/H]-[Mg/Fe] grid. We then used our empirical findings in combination with the flexible one-zone Chempy GCE model in order to fit the $\alpha_{\mathrm{IMF}}$, log$_{10}$(SNIa), log$_{10}$(SFE), and log$_{10}$(SFR$_{\mathrm{peak}}$) GCE model parameters and examined how these vary with [Fe/H], [Mg/Fe], age, ($R_{\mathrm{apo}}+R_{\mathrm{peri}}$)/2, and 
$z_{\mathrm{max}}$. Below we list the main conclusions obtained in this paper:

\begin{itemize}
\item We measure the age-[Ba/Fe] and age-[Eu/Fe] profiles of Galactic disc populations within a grid of twenty-five [Fe/H]-[Mg/Fe] chemical cells (only showing those cells containing more than 20 stars). We find that there is a clear trend of [Ba/Fe] with age, where young (old) stellar populations within a chemical cell have higher (lower) [Ba/Fe] abundances. This result corroborates findings from previous work focused on the entire Galactic disc (\citealp[e.g.,][]{Bedell2018,Magrini2018,Casamiquela2021, Zinn2021}), and implies that the genesis source of Ba is primarily produced in a nucleosynthetic channel that is independent of SN. Conversely, we find that the age-[Eu/Fe] profiles across chemical cells are flat. This suggests that Eu has a stronger dependence on SN contributions, for disc stars [Fe/H] $>$ $-1.0$ dex.
\item We find that both the age-[Ba/Fe] and age-[Eu/Fe] relations vary across the grid of chemical cells. Specifically, we find that for fixed [Mg/Fe], the age-[Ba/Fe] relations become flatter with increasing [Fe/H], and that for fixed [Fe/H], these relations increase in amplitude (at all ages) with decreasing [Mg/Fe]. Along similar lines, we show that the [Eu/Fe]-age relations decrease in amplitude (at all ages) with increasing [Fe/H]. Further, for fixed [Fe/H] these abundance-age relations increase in amplitude with increasing [Mg/Fe]. 
\item We note that at the oldest ages (namely, >10 Gyr), the age-[Ba/Fe] relations present an upturn that suggests an increase in [Ba/Fe] across all chemical cells. We note that this upturn could be explained in two ways: i) the onset of AGB-winds (that contribute Ba) has a slighly longer delay time than SNIa (that contribute primarily Fe). This would result in the primordial [Ba/Fe] ratio being diluted with SNIa explosions until Ba is synthetised in the winds of AGB stars, leading to an upturn at the oldest ages; ii) GALAH isn't able to reliable determine [Ba/Fe] for the oldest RGB stars (see also \citet[][]{Hayden2020}). Moreover, we also point out the increasing [Eu/Fe] abundance for the older stellar populations in the most [Fe/H]-poor and [Mg/Fe]-poor chemical cells, and suggest it is likely due for contamination in our sample from accreted populations that have been shown to present high [Eu/Fe] abundances.

\item We fit the mean [Ba/Fe], [Fe/H], [Mg/Fe], and age measurements for a total of twenty-seven different stellar populations contained within nine chemical cells that span a range in [Fe/H]-[Mg/Fe] space of --0.7 < [Fe/H] < 0.2 and --0.2 < [Mg/Fe] < 0.4, and span an age range of 2 < Age < 10 Gyr (where most of our data is contained and for which we obtain the most reliable [Ba/Fe] abundances), using the flexible one-zone Chempy GCE model. In doing so, we have let the $\alpha_{\mathrm{IMF}}$, log$_{10}$(SNIa), log$_{10}$(SFE), and log$_{10}$(SFR$_{\mathrm{peak}}$) model parameters be free, and have set out to find the optimised parameter values that best describe the data. Given the results from these fits, we find that the one-zone model adopted by Chempy is able to fit the data in the majority of cases, but suggests that the GCE model parameters optimised vary across the dimensions of age, [Mg/Fe], and [Fe/H] (see Fig~\ref{chempy_feh} and Fig~\ref{chempy_mgfe}). 
\item As the model predicts that the parameters vary with [Fe/H], [Mg/Fe], and age, and these measured quantities have been recently shown to be good proxy's for the birth radius of stars in the disc (\citealp[e.g.,][]{,Ness2019,Buck2020,Ness2021,Yuxi2021}), we explore the dependence of the optimised GCE model parameters with the mean Galactocentric position of the stellar populations fitted. More specifically, we explore the relation between mean orbital radius for the stellar populations fitted (Fig~\ref{chempy_rmean_fixedfeh}, but see also the results for when examining the mean maximum vertical height above the plane of the Galaxy (Fig~\ref{chempy_zs}). We find that these two measured quantities also correlate with the optimised GCE model parameters, and find one striking result that comes out from this exercise (see the left panel of Fig~\ref{chempy_rmean_fixedfeh}): stellar populations at $\sim$7 kpc are better fit by a GCE model with a more top-heavy high-mass slope of the IMF ($\alpha_{\mathrm{IMF}}$ $\sim$ --2.2), whereas those stellar populations at a larger Galactocentric radius ($\sim$10 kpc) prefer a more bottom-heavy IMF ($\alpha_{\mathrm{IMF}}$ $\sim$ --2.7). This result is more pronounced for the metal-poor populations (i.e., [Fe/H] < --0.1, but it also observed for the more metal-rich ones), and suggests that there is possibly a spatially (radially) varying IMF in the disc of the Galaxy. We note that this variation is only observed for the high-mass slope of this key GCE parameter.
\item We explore the dependence of orbital radius in the measured age-[Ba/Fe] relations in Fig~\ref{bafe_cells_rmean}. We see that at fixed age, the mean orbital radius value changes with varying [Mg/Fe] and [Fe/H]. At high [Mg/Fe] $>$ 0.2, [Fe/H] and [Mg/Fe] alone are indicative of the present day radius (the [Ba/Fe]-radius relations are flat). However, at [Mg/Fe] $<$ 0.2 dex, there is a strong dependence of the mean orbital radius within a cell, on the [Ba/Fe], and correspondingly, age of the stars. Here, stellar populations that reside in a neighbourhood of ([Mg/Fe],[Fe/H]) space and have higher [Ba/Fe] values (and younger ages) have, on average, a larger mean orbital radius than those populations with lower [Ba/Fe] abundances (and are older).  We suggest that these correlations arise because (Fe,Mg,age), and additionally, Ba for [Mg/Fe] $<$ 0.2 dex stars, are a link to birth radius, and we are seeing an imprint of the  correlation between current day radius and birth radius - traced by abundances and age -  in this plane (\citealp[e.g.,][]{Ness2019,Buck2020}).
\end{itemize}

This paper attempts to frame empirical abundance-age relations of neutron capture elements across a novel supernovae reference frame via Galactic chemical evolution modelling. Using the assumptions of a flexible single-zone chemical evolution model, we have been able to successfully interpret our observational results in a theoretical setting. Future works aiming at modelling the chemical evolution of the Galactic disc will require more sophisticated multi-zone GCE models that also enable the flexibility of Chempy, allowing the optimization of the model parameters.

\section*{Acknowledgements}
The authors thank the referee for the thoughtful review that improved the clarity of the manuscript. DH thanks Sue, Alex, and Debra for their constant love and support. He also thanks Oliver Philcox for help with the handling of the Chempy software, Joel Zinn for helpful comments on the earlier versions of the manuscript, and the astronomy department at the University of Queensland for being so kindly welcoming and allowing him to work at the university during his visit in Australia. DH also warmly thanks Emily Cunningham for her support over the past year, and for help in establishing collaborations that led to this work. MKN is in part supported by a Sloan Foundation Fellowship. JR was funded by the DLR (German space agency) via grant 50QG1403.

This work uses data from the GALAH Survey, acquired through the Australian Astronomical Observatory, under programs: A/2013B/13 (The GALAH
pilot survey); A/2014A/25, A/2015A/19, A2017A/18 (The GALAH survey phase 1), A2018 A/18 (Open clusters with HERMES), A2019A/1 (Hierarchical star formation in Ori OB1), A2019A/15
(The GALAH survey phase 2), A/2015B/19, A/2016A/22,
A/2016B/10, A/2017B/16, A/2018B/15 (The HERMES-TESS program), and A/2015A/3, A/2015B/1, A/2015B/19, A/2016A/22, A/2016B/12, A/2017A/14, (The HERMES K2-follow-up program).
This paper includes data that has been provided by AAO Data Central (datacentral.aao.gov.au). This work has also made use of data from the European Space Agency (ESA) mission \textit{Gaia} (\href{http://www.cosmos.esa.int/gaia}{http://www.cosmos.esa.int/gaia}), processed by the \textit{Gaia} Data Processing and Analysis Consortium (DPAC, \href{http://www.cosmos.esa.int/web/gaia/dpac/consortium}{http://www.cosmos.esa.int/web/gaia/dpac/consortium}). Funding for the DPAC has been provided by national institutions, in particular the institutions participating in the \textit{Gaia} Multilateral Agreement. Other facilities: This publication makes use of data products from the Two
Micron All Sky Survey \citep[][]{Skrutskie2006} and the CDS VizieR
catalogue access tool \citep[][]{Ochsenbein2000}.

{\it Software:} Astropy \citep{astropy:2013,astropy:2018}, Chempy: \citep{Rybizki2017}, NumPy \citep{NumPy}, Matplotlib \citep{Hunter:2007},
Galpy \citep{Galpy2015,Galpy2018}.

\section*{Data availability}
The data used for this study can
be accessed publicly via \href{https://docs.datacentral.org.au/galah/dr3/overview/}{https://docs.datacentral.org.au/galah/dr3/overview/}.

\bibliographystyle{mnras}
\bibliography{refs}

\appendix

\section{Further results from Chempy modelling}

\begin{figure*}
    \centering
    \includegraphics[width=\textwidth]{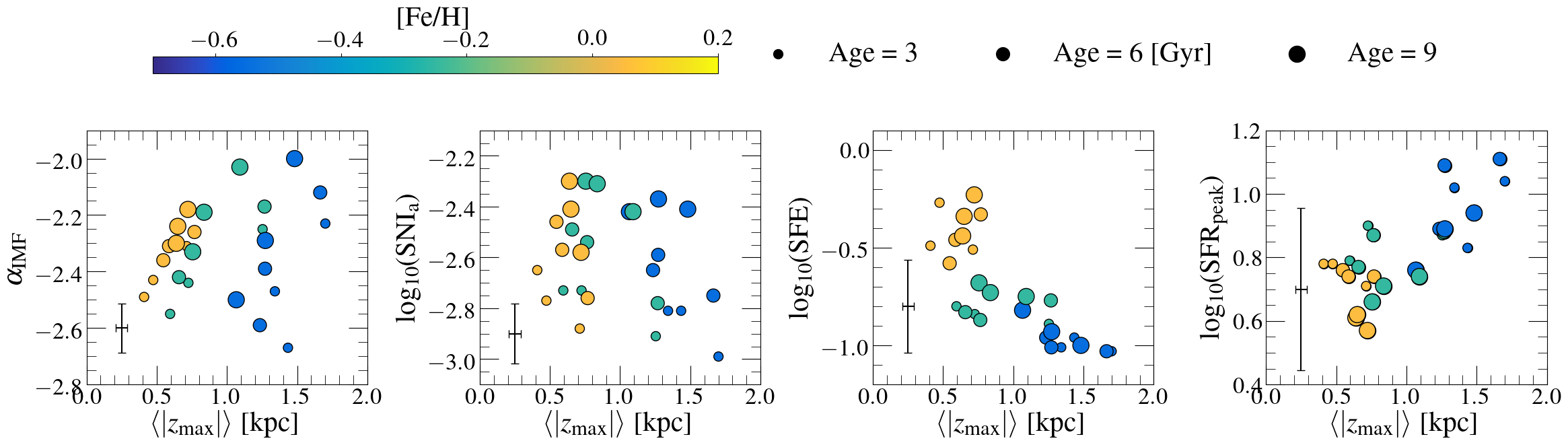}
    \caption{Same as Fig~\ref{chempy_rmean_fixedfeh}, but now as a function of mean maximum vertical excursion above the Galactic plane. As for orbital radius, we also see a link between the maximum position of a star above the Galactic disc and the model parameters estimated.}
    \label{chempy_zs}
\end{figure*}

\begin{figure*}
    \centering
    \includegraphics[width=\textwidth]{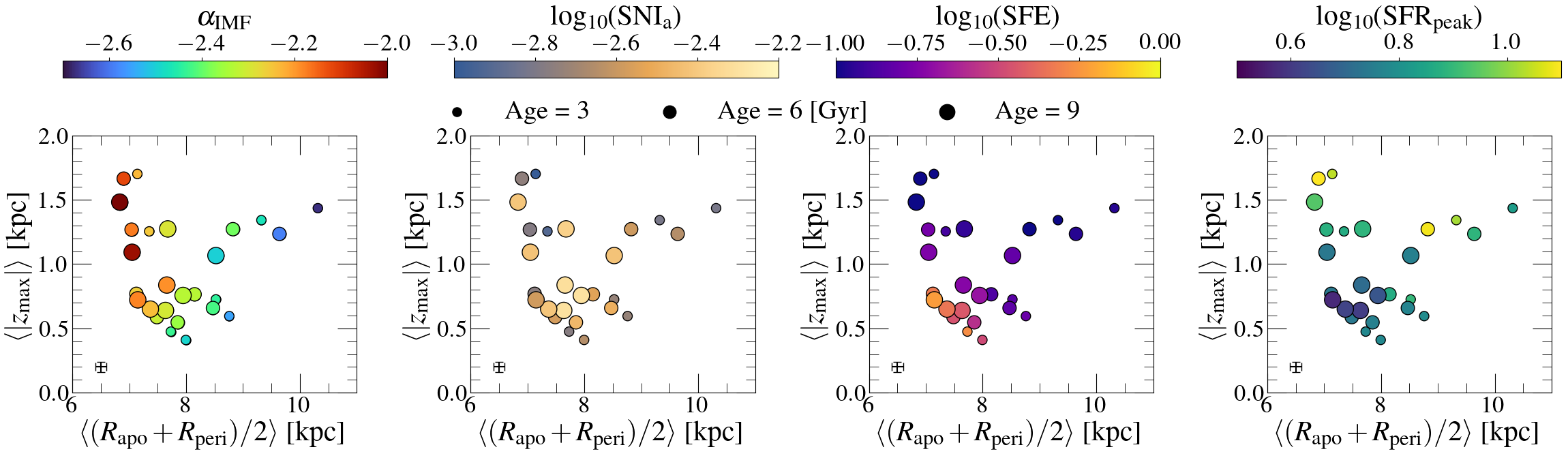}
    \caption{Mean vertical height vs mean orbital radius for the twenty-seven stellar populations fitted with Chempy in the nine chemical cells containing the majority of our sample. Each plot is colour coded by the resulting best fit parameters from the Chempy GCE models.}
    \label{chempy_rmean_zmax}
\end{figure*}

\begin{figure*}
    \centering
    \includegraphics[width=\textwidth]{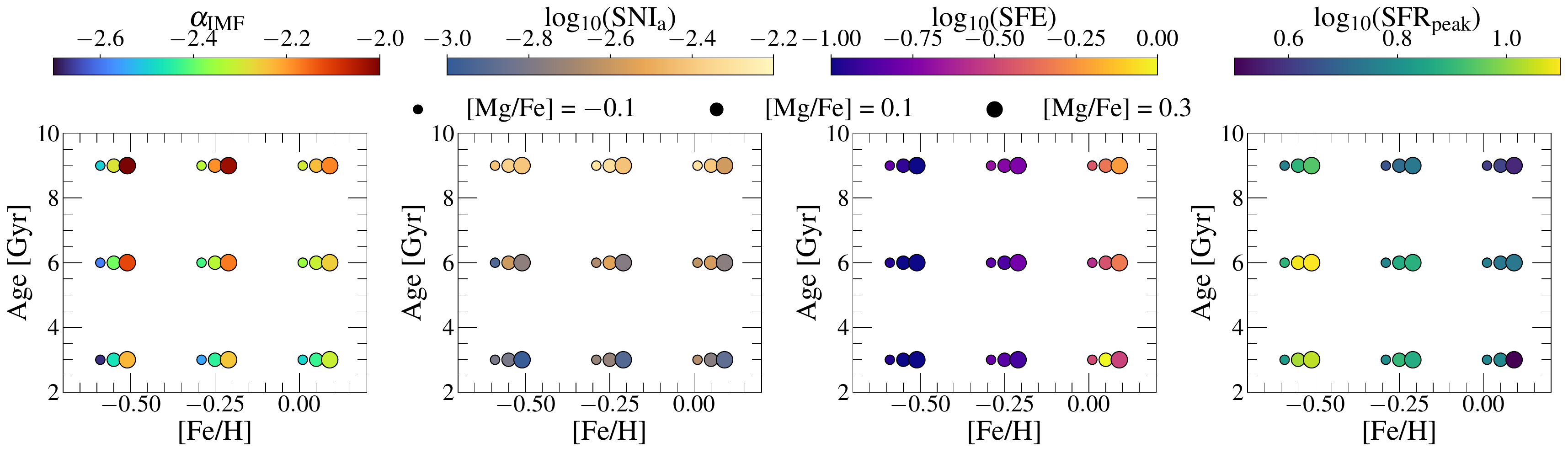}
    \caption{Summary of the results determined from fitting the twenty-seven different stellar populations (within nine chemical cells) with Chempy. Here the [Fe/H] and age values are approximate values to the true age and [Fe/H] values for the stellar populations plotted (used for illustrative purposes). The model parameters vary smoothly across the three pivot parameters: [Fe/H], [Mg/Fe], and age.}
    \label{chempy_summary}
\end{figure*}

\section{$\sigma_{\mathrm{[X/Fe]}}$-age relations}
\label{app_scatters}

\begin{figure*}
    \centering
    \includegraphics[width=\textwidth]{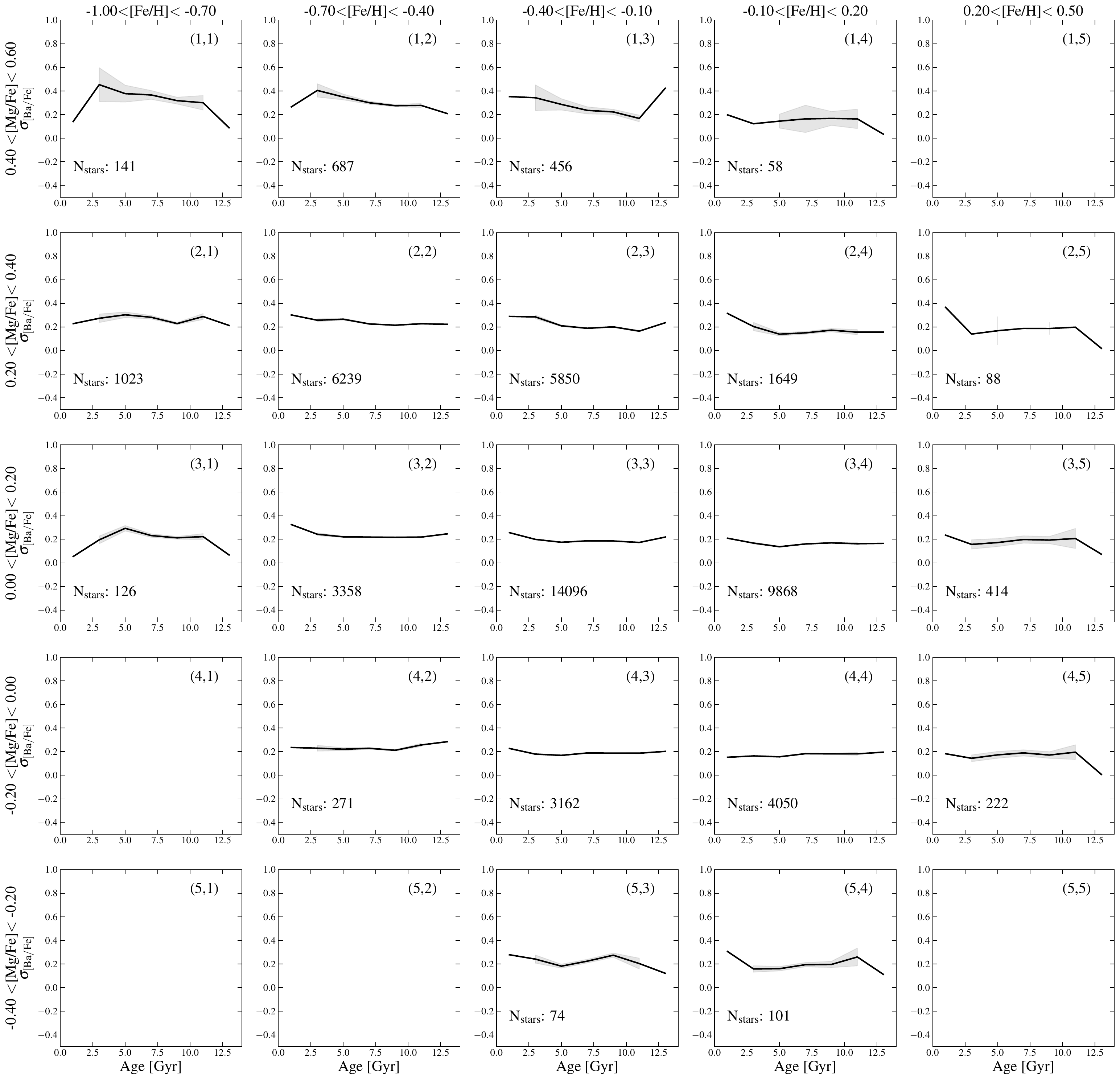}
    \caption{Same as Fig~\ref{bafe_cells} but for the 1-$\sigma$ dispersion values in the chemical cells.}
    \label{bafe_cells_uncer}
\end{figure*}

\begin{figure*}
    \centering
    \includegraphics[width=\textwidth]{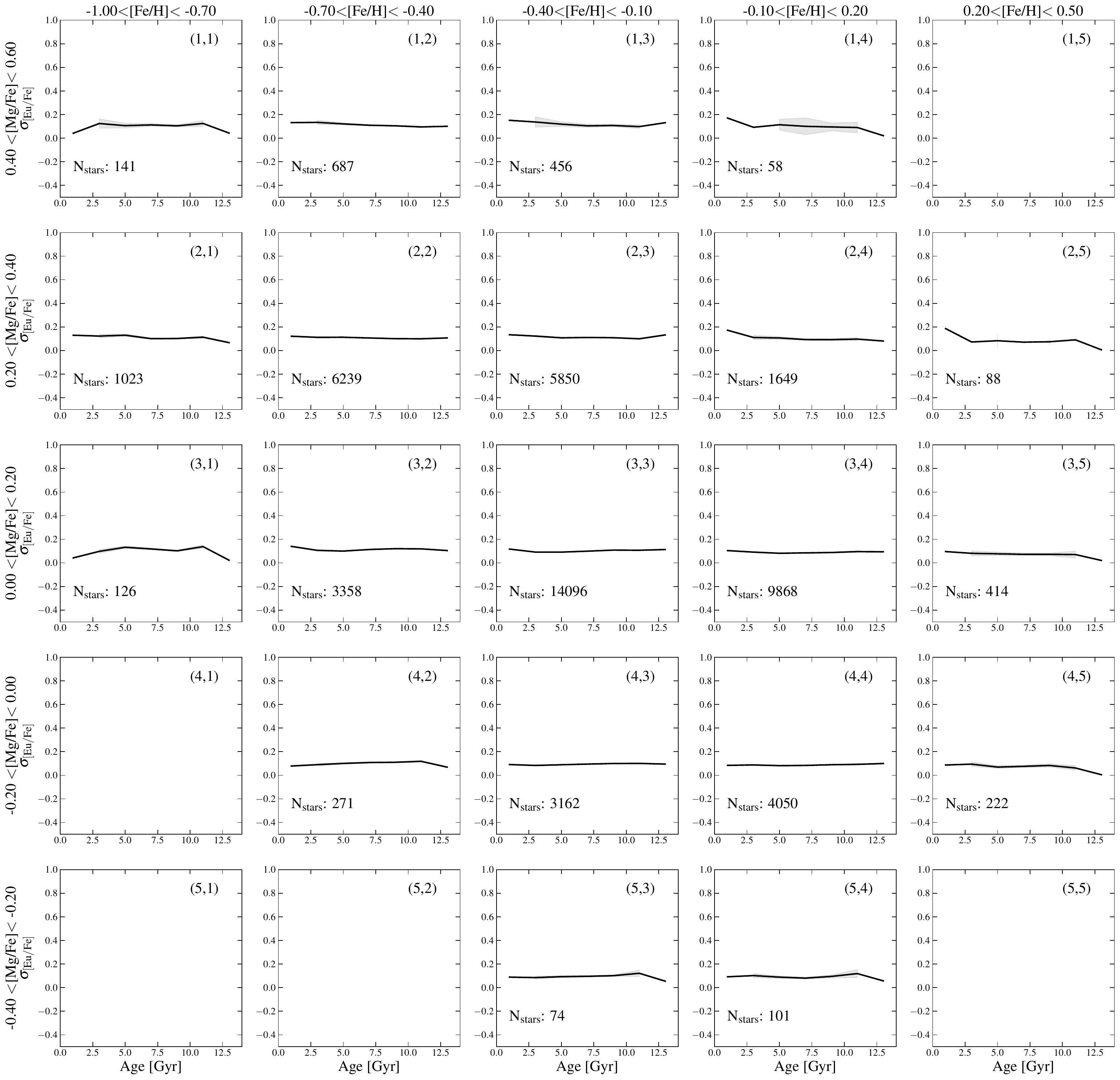}
    \caption{Same as Fig~\ref{bafe_cells_uncer} but for [Eu/Fe].}
    \label{eufe_cells_uncer}
\end{figure*}

\section{[Ba/Fe]-orbital radius relations}
\label{app_bafe_rmean}
\begin{figure*}
    \centering
    \includegraphics[width=\textwidth]{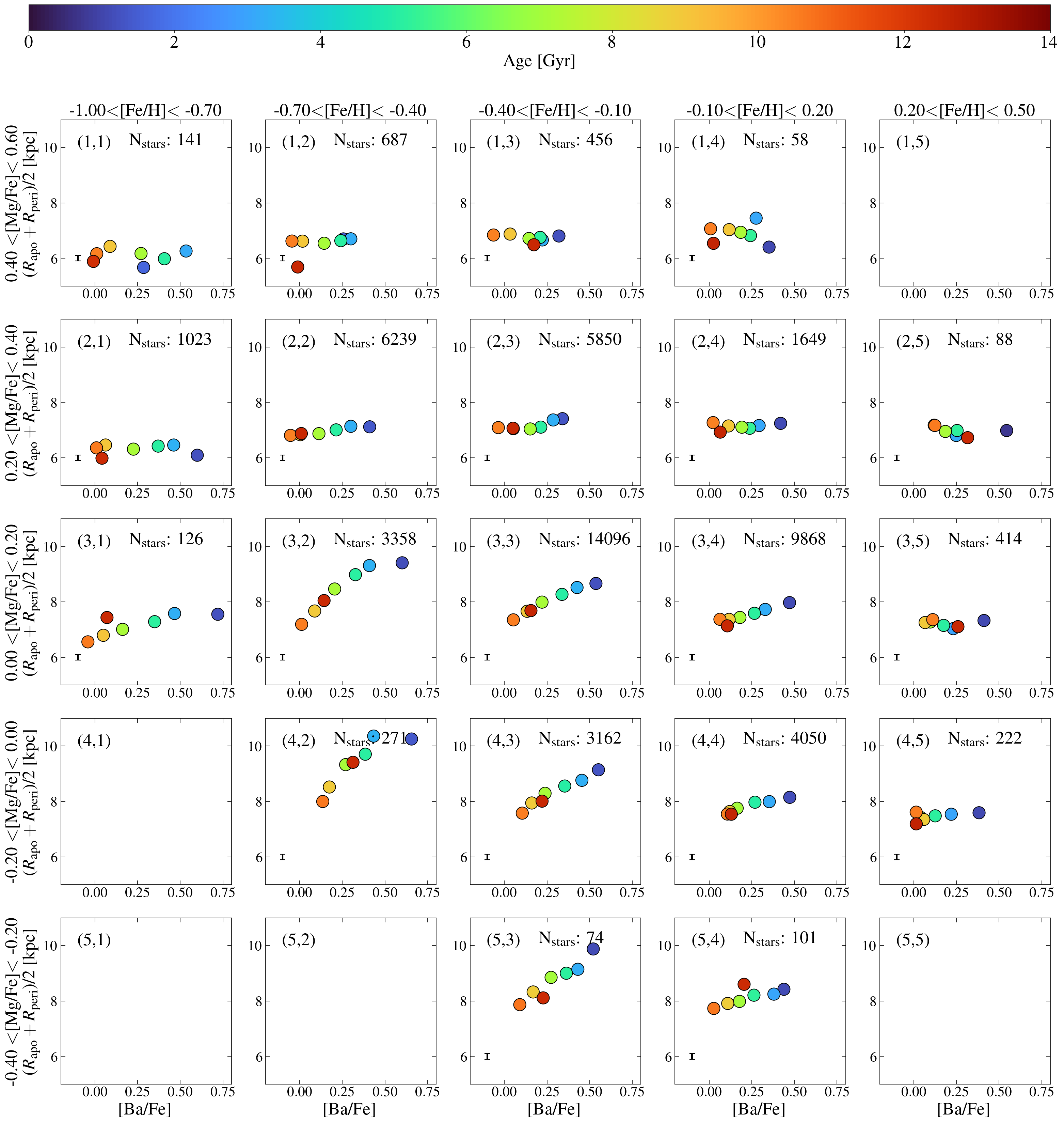}
    \caption{5$\times$5 grid of chemical cells as in Fig~\ref{bafe_cells}, but now plotting the mean orbital radius as a function of the mean [Ba/Fe] for every stellar population in each cell binned by age (i.e., the colour bar). There is a clear relation between orbital radius and [Ba/Fe] for the [Mg/Fe] < 0.2 cells, which becomes flat for the more [Mg/Fe]-rich cells. }
    \label{bafe_rmean_age}
\end{figure*}

\section{Assessing how well Chempy models the chemical abundance data in chemical cells}
\label{app_rms_cells}

\begin{figure*}
    \centering
    \includegraphics[width=\textwidth]{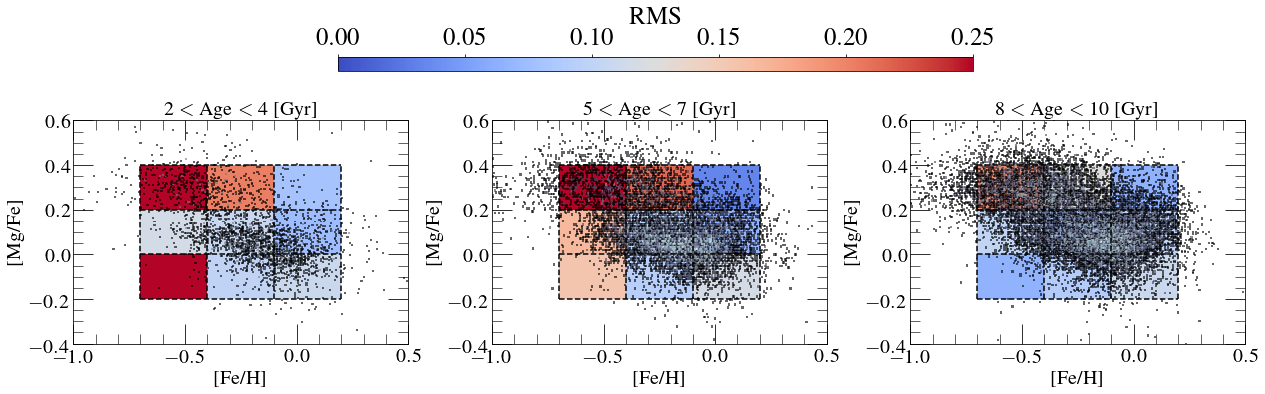}
    \caption{2D density distribution of the parent sample binned by age and gridded into the nine chemical cells we model using the flexible Chempy GCE model. Each cell is colour coded by the root-mean-squared (RMS) value, computed as: RMS = $\sqrt{(\mathrm{[Fe/H]_{obs} - \mathrm{[Fe/H]_{pred}})^{2} + (\mathrm{[Mg/Fe]_{obs}} - \mathrm{[Mg/Fe]_{pred}})^{2} + (\mathrm{[Ba/Fe]_{obs}} - \mathrm{[Ba/Fe]_{pred}})^{2}}}$. A low(high) RMS value signifies a better(worse) fit of the model to the data. All our chemical cell fits are relatively well modelled within the 1-$\sigma$ uncertainties. However, we find that on average, those more metal-poor cells (i.e., --0.7 < [Fe/H] < --0.4) are less well fitted by the Chempy model.  }
    \label{rms_cells}
\end{figure*}

\section{Samples from the Monte-Carlo Markov-Chain routine on the chemical cells}
\label{app_mcmc}

\begin{figure}
    \centering
    \includegraphics[width=\columnwidth]{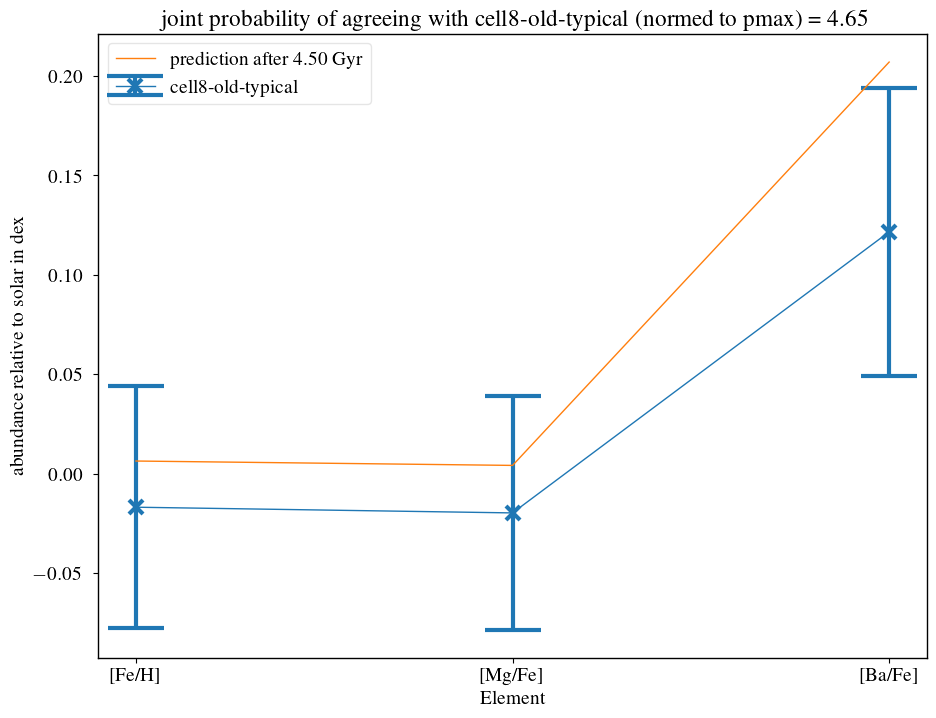}
    \caption{Resulting abundance predictions from Chempy compared to the observed abundances for the old stellar population in cell (4,4).}
    \label{samples_cell13_old}
\end{figure}

\begin{figure}
    \centering
    \includegraphics[width=\columnwidth]{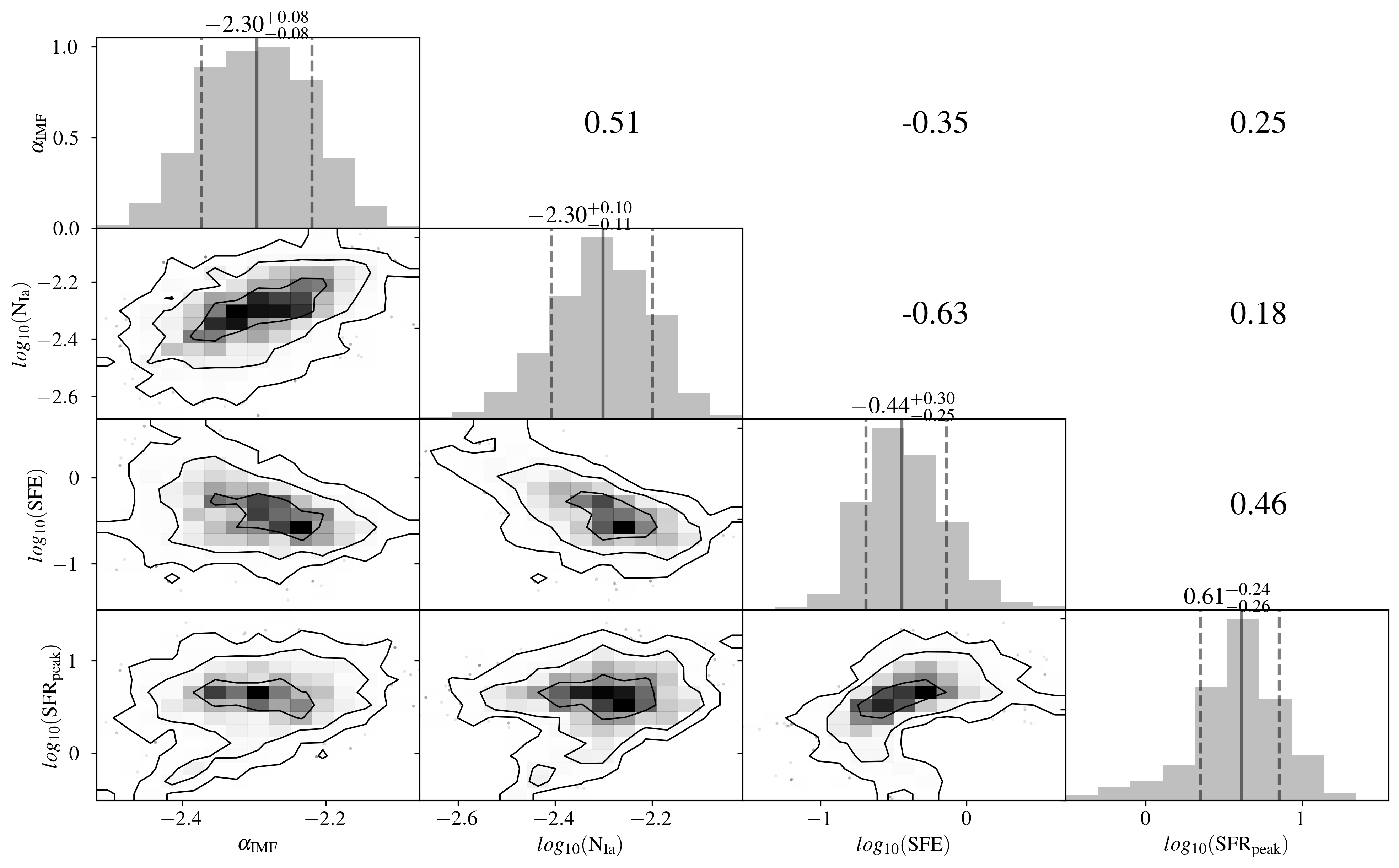}
    \caption{Resulting posterior distribution of the model parameters for the old population in cell (4,4) outputted from Chempy. The numbers on top of each parameter are the correlation values outputted from Chempy between each of the model parameters.}
    \label{samples_cell13_old_samples}
\end{figure}
\begin{figure}
    \centering
    \includegraphics[width=\columnwidth]{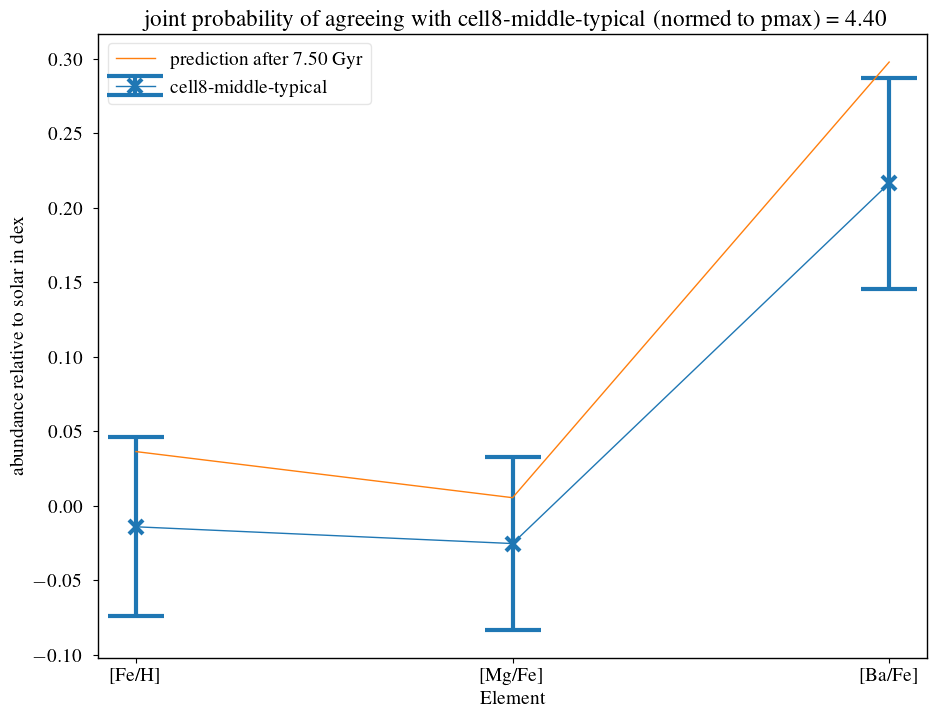}
    \caption{Same as Fig~\ref{samples_cell13_old} but for the intermediate age bin.}
    \label{final_cell13_middle}
\end{figure}

\begin{figure}
    \centering
    \includegraphics[width=\columnwidth]{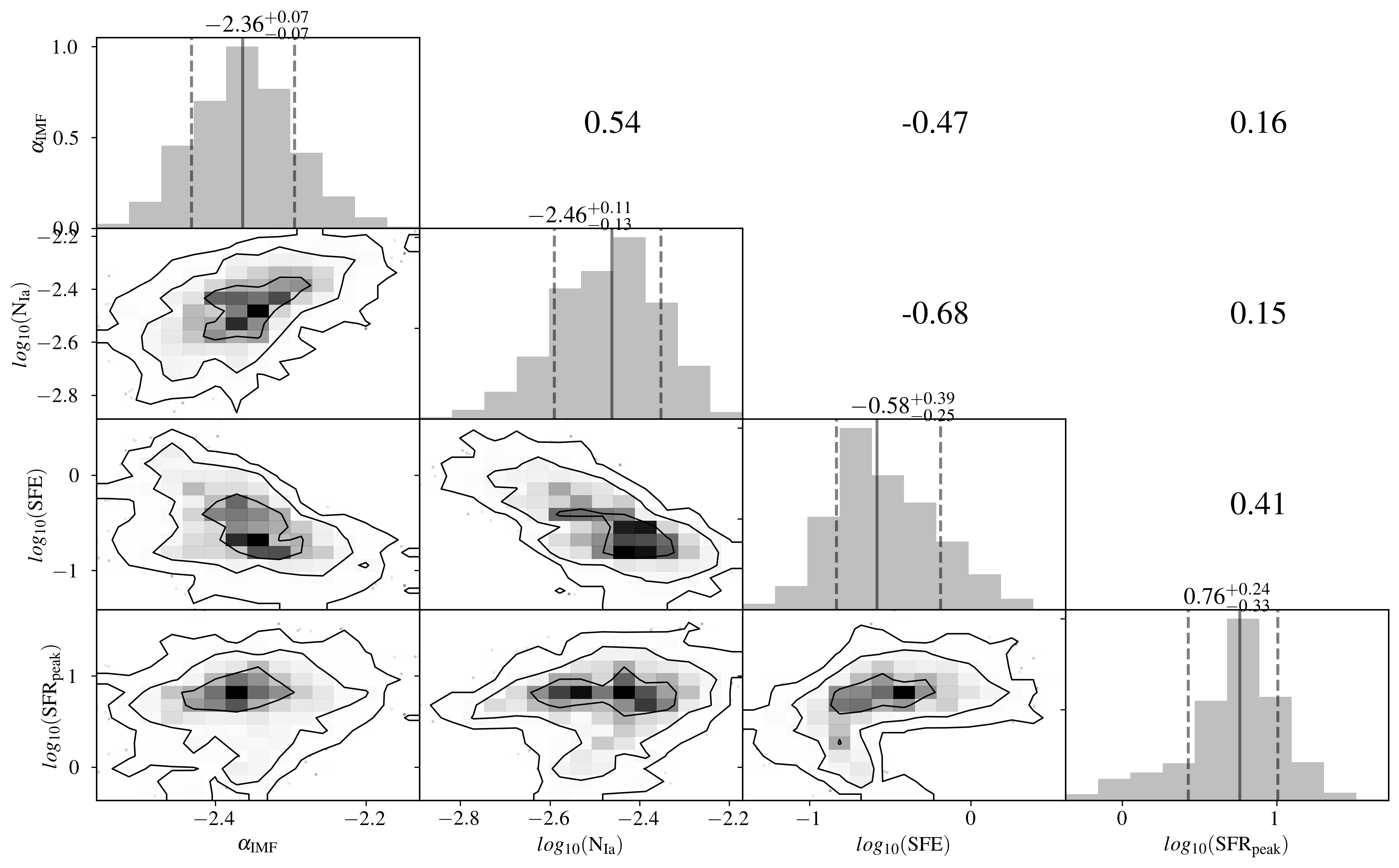}
    \caption{Same as Fig~\ref{samples_cell13_old_samples} but for the intermediate age bin.}
    \label{final_cell13_middle_samples}
\end{figure}

\begin{figure}
    \centering
    \includegraphics[width=\columnwidth]{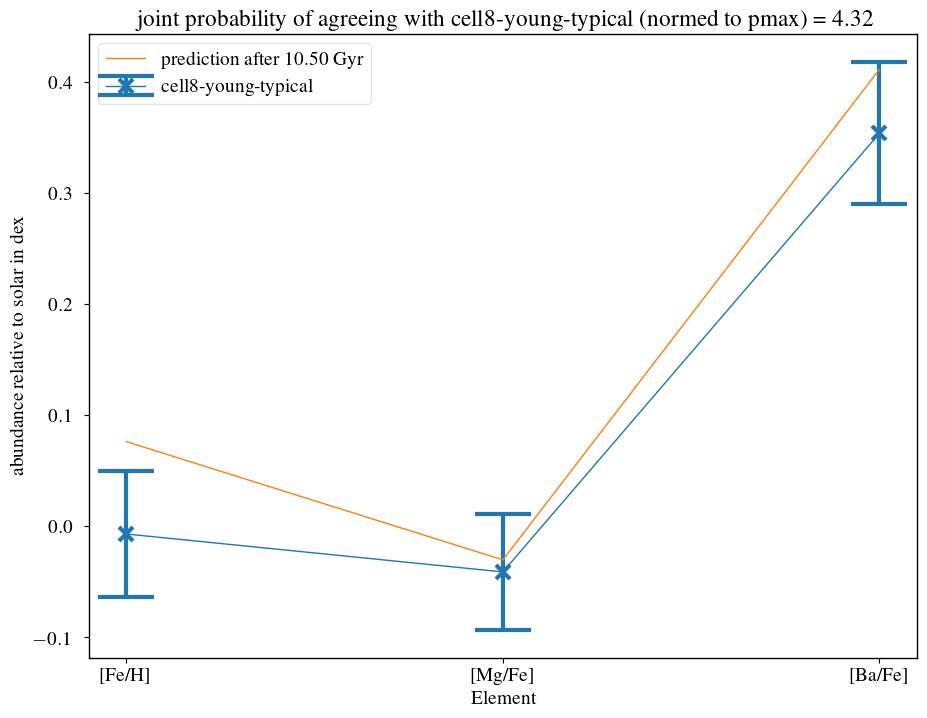}
    \caption{Same as Fig~\ref{samples_cell13_old} but for the young age bin.}
    \label{samples_cell13_young}
\end{figure}

\begin{figure}
    \centering
    \includegraphics[width=\columnwidth]{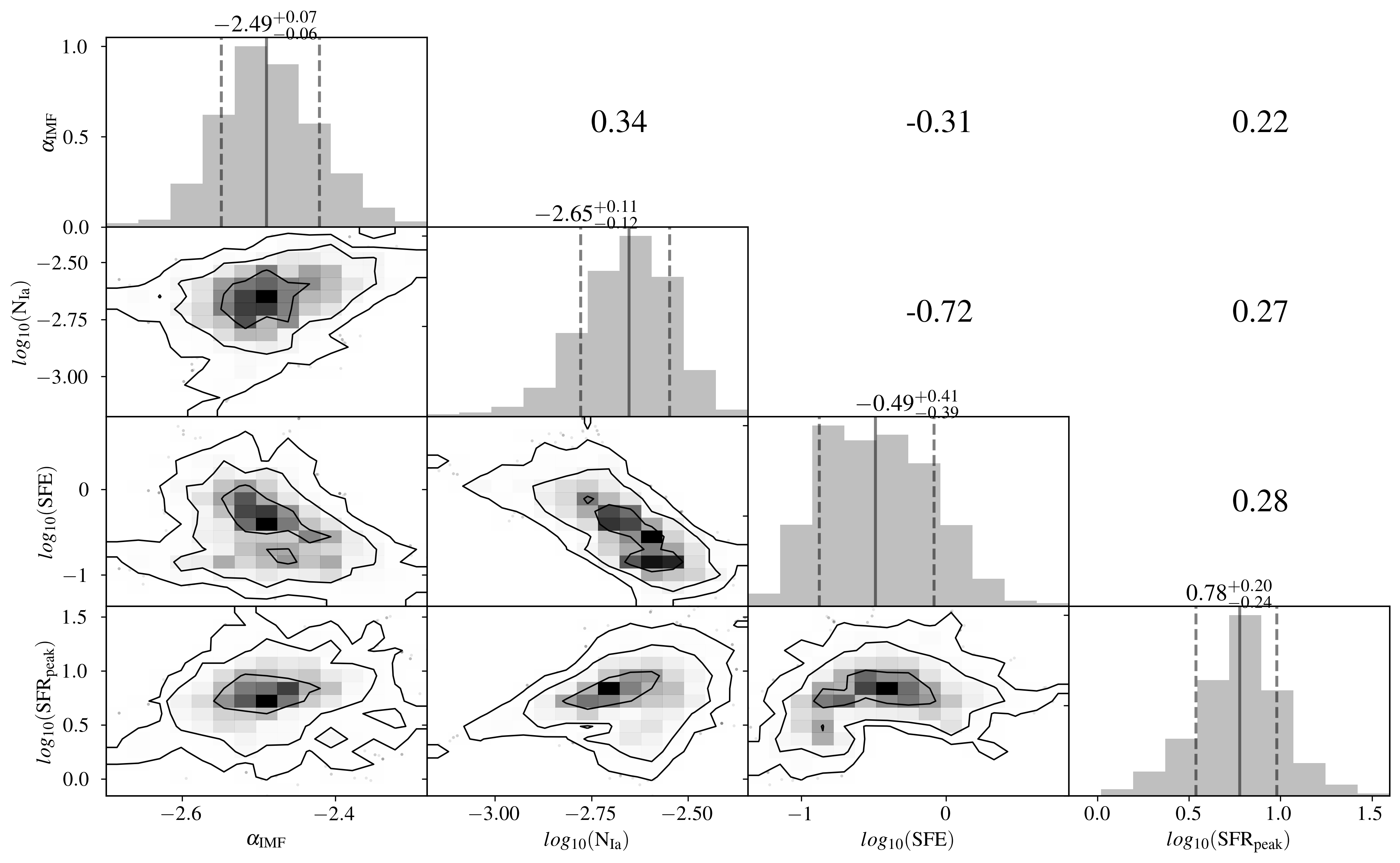}
    \caption{Same as Fig~\ref{samples_cell13_old_samples} but for the young age bin.}
    \label{samples_cell13_young_samples}
\end{figure}


\bsp	
\label{lastpage}
\end{document}